\documentclass{emulateapj}
\usepackage{color} 
\usepackage[colorlinks,urlcolor=blue,citecolor=blue,linkcolor=blue]{hyper ref}

\usepackage{epsfig}
\usepackage{subfigure}
\usepackage{graphicx}
\usepackage{amsmath}
\usepackage{natbib}
\newcommand{\pa}{\partial}
\newcommand{\mb}{\boldsymbol}

\shorttitle{Magnetic Flux Transport in PPDs}
\shortauthors{Bai \& Stone}

\begin{document}


\title{Hall-effect Mediated Magnetic Flux Transport in Protoplanetary Disks}


\author{Xue-Ning Bai\altaffilmark{1} and James M. Stone\altaffilmark{2}}
\affil{$^1$Institute for Theory and Computation, Harvard-Smithsonian
Center for Astrophysics, 60 Garden St., MS-51, Cambridge, MA 02138}
\affil{$^2$Department of Astrophysical Sciences, Peyton Hall, Princeton
University, Princeton, NJ 08544}
\email{xbai@cfa.harvard.edu}




\begin{abstract}
The global evolution of protoplanetary disks (PPDs) has recently been shown to be largely
controlled by the amount of poloidal magnetic flux threading the disk. The amount of
magnetic flux must also co-evolve with the disk, as a result of magnetic flux transport, a process
which is poorly understood. In weakly ionized gas as in PPDs, magnetic flux is largely frozen
in the electron fluid, except when resistivity is large. When the disk is largely laminar, we
show that the relative drift between the electrons and ions (the Hall-drift), and the ions and
neutral fluids (ambipolar-drift) can play a dominant role on the transport of magnetic flux.
Using two-dimensional simulations that incorporate the Hall effect and ambipolar diffusion
(AD) with prescribed diffusivities, we show that when large-scale poloidal field is aligned
with disk rotation, the Hall effect rapidly drags magnetic flux inward at the midplane region,
while it slowly pushes flux outward above/below the midplane. This leads to a highly radially
elongated field configuration as a global manifestation of the Hall-shear instability. This field
configuration further promotes rapid outward flux transport by AD at the midplane, leading
to instability saturation. In quasi-steady state, magnetic flux is transported outward at
approximately the same rate at all heights, and the rate is comparable to the Hall-free case.
For anti-aligned field polarity, the Hall effect consistently transports magnetic flux outward,
leading to a largely vertical field configuration in the midplane region. The field lines in the
upper layer first bend radially inward and then outward to launch a disk wind.
Overall, the net rate of outward flux transport is about twice faster than the aligned case.
In addition, the rate of flux transport increases with increasing disk magnetization.
The absolute rate of transport is sensitive to disk microphysics which remains to be explored
in future studies.
\end{abstract}



\section{Introduction}\label{sec:intro}

Global structure and evolution of protoplanetary disks (PPDs) play a fundamental
role in almost all stages of planet formation. It has recently been realized that due
to the weakly ionized nature of PPD gas, the magnetorotational instability (MRI,
\citealp{BH91}) is almost entirely suppressed in the inner region of PPDs
\citep{BaiStone13b,Bai13,Gressel_etal15}, and is substantially damped in the outer
disk \citep{Simon_etal13b,Bai15}. Efficient angular momentum transport requires the
disk to be threaded with external large-scale poloidal magnetic flux, presumably
inherited from the star formation process, and a magnetized disk wind is likely the
primary mechanism to drive disk accretion. Because the wind kinematics strongly
depends on disk magnetization (e.g., \citealp{Bai_etal16}), global disk evolution is
primarily governed by the amount of poloidal magnetic flux threading the disks, and
its radial distribution \citep{Bai16}.
Before we can fully understand global disk evolution, a more fundamental question
is, what determines the amount and distribution of magnetic flux threading PPDs?
Equivalently, how is magnetic flux transported in PPDs?

Magnetic flux transport has conventionally been modeled as a competition between
inward advection by viscously-driven accretion, and outward diffusion by (turbulent or
physical) resistivity \citep{Lubow_etal94a}. While more recent works have taken into
account disk vertical structure \citep{RothsteinLovelace08,GuiletOgilvie12,GuiletOgilvie13},
or radial resistivity profile \citep{Okuzumi_etal14,TakeuchiOkuzumi14}, they all fall
into the same advection-diffusion framework, which ignores the wind-driven accretion
process, and detailed disk microphysics.

Weakly ionized PPDs are subject to three non-ideal magnetohydrodynamic (MHD)
effects, namely, Ohmic resistivity, the Hall effect, and ambipolar diffusion (AD). To our
knowledge, the Hall effect and AD have not been considered in the theory of magnetic
flux transport in accretion disks.\footnote{By contrast, the role of non-ideal MHD effects,
especially AD, on the ``magnetic flux problem" in star formation has been studied
extensively in the literature (see \citet{McKeeOstriker07,Li_etal14} for reviews), and is
still undergoing active develoopment.}
The problem of magnetic flux transport is directly coupled to the gas dynamics, and
current studies in PPDs have mostly focused on the gas dynamics itself.

Unlike resistivity, which diffuses magnetic field isotropically, both AD and the Hall
effect are anisotropic. AD shares some similarities with Ohmic resistivity that acts to
diffuse magnetic flux outwards for typical field configurations, its
anisotropic nature also introduces novel ingredients, as we will address in this paper.
The Hall effect behaves completely differently, which we will focus on in this work.
It is well known that the Hall term affects the disk dynamics in polarity-dependent
ways \citep{Wardle99,WardleSalmeron12}. Local semi-analytical wind solutions that
include these effects have been studied in the literature (e.g.,
\citealp{WardleKoenigl93,Konigl_etal10,Salmeron_etal11}). Local shearing-box
simulations with imposed constant net vertical field have also found very different
behaviors for different field polarities
\citep{SanoStone02a,SanoStone02b,KunzLesur13,Bai14,Bai15,Simon_etal15b}.
Evidence for rapid and polarity-dependent magnetic flux transport has been reported
\citep{Bai14}. However, being local solutions/simulations, the results depend on the
imposed boundary conditions. Consequently, the rate of magnetic flux transport can
not be reliably determined, and in some semi-analytical solutions, it is in fact
imposed as a free parameter.

In this paper, we first point out in Section \ref{sec:physics} that the Hall effect affects
magnetic flux transport in PPDs in a dramatic way depending on the polarity of
poloidal field threading the disk. We describe in Section \ref{sec:setup} a set of
two-dimensional (2D) global disk simulations that incorporate both the Hall effect
and AD to study magnetic flux transport in PPDs. These simulations have simple
prescriptions of disk ionization and thermodynamics, and can be considered
as controlled experiments aiming to demonstrate and clarify the basic physics of
magnetic flux transport in a {\it laminar} disk. Main results on flux transport are
presented in Section \ref{sec:overview}, which confirm our theoretical expectations.
We discuss the gas dynamics in the simulations in Section \ref{sec:gasdyn} and
analyze flux transport in more detail in Section \ref{sec:Btrans}. In Section
\ref{sec:beta}, we discuss how the rate of flux transport depends on disk
magnetization. Implications and limitations of the results are discussed in Section
\ref{sec:discussion}. In Section \ref{sec:sum}, we summarize and conclude.

\section[]{Basic Physics}\label{sec:physics}

In a weakly ionized gas, magnetic fields are no longer frozen to the bulk gas
(i.e., neutrals), but are effectively carried by tracer amount of ionized species.
The physics is most transparently explained when electrons and ions are the
only ionized species (i.e., ignoring charged dust grains), as we assume here.
Being the most mobile species, magnetic flux is largely frozen into the electrons,
modulo the effect of electron-neutral collisions (Ohmic resistivity) which allows
magnetic field to slide through the electron fluid. However, electrons do not
necessarily move with the bulk gas, and we can decompose the electron velocity
${\mb v}_e$ into
\begin{equation}\label{eq:ve}
{\mb v}_e={\mb v}+({\mb v}_e-{\mb v}_i)+({\mb v}_i-{\mb v})\ ,
\end{equation}
where ${\mb v}$ and ${\mb v}_i$ are the velocities of the bulk gas (neutrals), and
the ions. The electron-ion drift ${\mb v}_e-{\mb v}_i$, also known as the {\it Hall-drift},
corresponds to the Hall effect, and the ion-neutral drift ${\mb v}_i-{\mb v}$,
also known as {\it ambipolar drift}, corresponds to ambipolar diffusion (AD).

\begin{figure*}
    \centering
    \includegraphics[width=160mm]{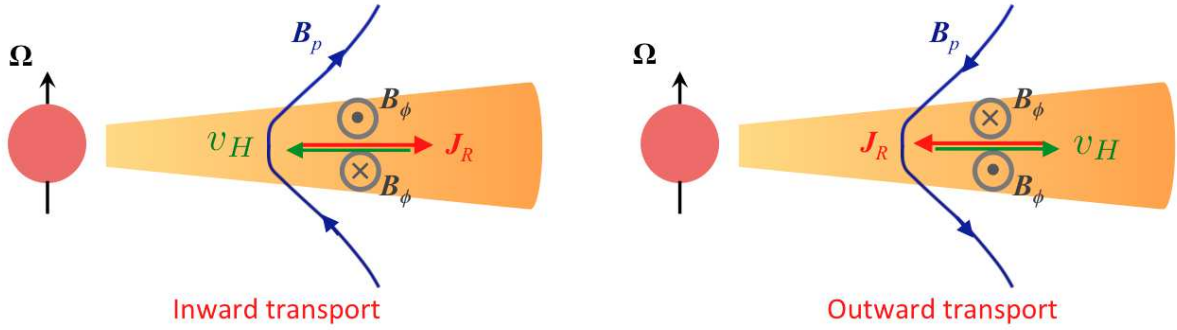}
  \caption{Illustration of the Hall-effect mediated magnetic flux transport.
  In the Hall dominated regime, magnetic flux is transported along the direction of the
  Hall-drift at speed $v_H$, which is opposite to the direction radial current $J_R$
  resulting mainly from the vertical toroidal field gradient. The direction of transport
  then depends on the polarity of the large-scale poloidal magnetic field
  relative to the disk rotation axis (marked with $\Omega$). Note that this Cartoon only
  illustrates magnetic flux transport in the midplane region. The situation elsewhere
  is different. See text for details.}\label{fig:cartoon}
\end{figure*}

The electron-ion drift is directly related to current density
${\mb J}=-en_e({\mb v}_e-{\mb v}_i)=(c/4\pi)\nabla\times{\mb B}$\ . In collisional
equilibrium (which is well satisfied in PPDs), the ion-neutral drift velocity is
determined by the balance between Lorentz force experienced by the ions and
the ion-neutral collisional drag
\begin{equation}
\gamma_i\rho\rho_i({\mb v}_i-{\mb v})=\frac{1}{c}{\mb J}\times{\mb B}\ ,
\end{equation}
where $\gamma_i$ is the coefficient of momentum transfer between ion-neutral
collisions, $\rho$, $\rho_i$ are the density of the bulk gas (neutrals) and the ions.
In particular, $\gamma_i\rho_i$ characterizes the frequency for the neutrals to
collide with the ions.

More formally, the evolution of magnetic field ${\mb B}$ is described by the
induction equation, given by
\begin{equation}
\frac{\pa{\mb B}}{\pa t}=\nabla\times({\mb v}_e\times{\mb B})
-\frac{4\pi}{c}\nabla\times(\eta_O{\mb J})\ ,\label{eq:indeq}
\end{equation}
where $\eta_O$ is Ohmic resistivity.
Substituting the electron velocity (\ref{eq:ve}) to the above,
one obtains the more familiar expression
\begin{equation}\label{eq:ind}
\frac{\pa{\mb B}}{\pa t}=\nabla\times({\mb v}\times{\mb B})-
\frac{4\pi}{c}\nabla\times(\eta_O{\mb J}+\eta_H{\mb J}\times{\mb b}+\eta_A{\mb J}_\perp)\ ,
\end{equation}
where ${\mb b}\equiv{\mb B}/B$ is the unit vector for the magnetic field direction,
${\mb J}_\perp=-({\mb J}\times{\mb b})\times{\mb b}$ is the component of
${\mb J}$ that is perpendicular to the magnetic field. The Hall and ambipolar diffusivities
(grain-free case) are given by
\begin{equation}
\eta_H=\frac{cB}{4\pi en_e}\equiv v_Al_H\ ,\quad
\eta_A=\frac{B^2}{4\pi\gamma_i\rho\rho_i}\equiv\frac{v_A^2}{Am\cdot\Omega}\ ,
\end{equation}
where we have defined the Hall length $l_H$ \citep{KunzLesur13}, which is the analog of
the ion inertial length in fully ionized plasmas, and the dimensionless AD Elsasser number
$Am\equiv v_A^2/\eta_A\Omega$, with $\Omega$ being
the disk Keplerian frequency. Both $l_H$ and $Am$ have the advantage of being independent
of magnetic field strength. The strength of the Hall term is also conveniently measured by the
dimensionless Hall Elsasser number defined by $\chi\equiv v_A^2/\eta_H\Omega$, which is
field-strength dependent. Note that at fixed ionization fraction $n_e/n$, we have
$\eta_O=$ constant, $\eta_H\propto B/\rho$, and $\eta_A\propto B^2/\rho^2$. Therefore,
AD becomes progressively more important towards lower density regions.

Using the Hall and AD diffusivities, one can further express the Hall and ambipolar drift velocites
as
\begin{equation}
{\mb v}_H\equiv{\mb v}_e-{\mb v}_i=-\frac{\eta_H}{B}(\frac{4\pi}{c}){\mb J}
=-v_A\frac{l_H\nabla\times{\mb B}}{B}\ ,\label{eq:vH}
\end{equation}
\begin{equation}
{\mb v}_{AD}\equiv{\mb v}_i-{\mb v}=\frac{\eta_A}{B}(\frac{4\pi}{c}){\mb J}\times{\mb b}
=\frac{(\nabla\times{\mb B})\times{\mb B}}{4\pi\rho\Omega\cdot Am}\ .\label{eq:vAD}
\end{equation}

From (\ref{eq:indeq}), transport of magnetic flux in a laminar flow is given by
\begin{equation}
\begin{split}\label{eq:Btrans}
\frac{d\Phi_B(R,z)}{dt}=&-2\pi R{\mathcal E}_\phi\\
=&-2\pi R\bigg[(v_{e,R}B_z-v_{e,z}B_R)+\frac{4\pi}{c}\eta_OJ_\phi\bigg]\ .
\end{split}
\end{equation}
where $\Phi_B(R,z)$ is the amount of magnetic flux enclosed within a ring at
cylindrical radius $R$ and height $z$, $\mathcal E$ is the electric field. In the
second equality, the first term corresponds to radial advection of vertical field that
directly lead to accumulation or reduction of the enclosed magnetic flux. The
second term corresponds to vertical advection of radial field into or out of the
ring, which also affects the amount of flux enclosed in the ring. As we will see,
both terms contribute to magnetic flux transport.

We focus on the Hall drift in this section. In thin disks, we generally expect
the toroidal field to be the dominant field component, and the current density
is mainly set by its vertical gradient. Therefore, we expect $J_R\gg J_z$,
and hence the Hall-effect mediated flux transport is dominated by the
radial advection term $v_{H,R}B_z$. Assuming axisymmetry, we have from
(\ref{eq:vH})
\begin{equation}
v_{H,R}=\frac{\eta_H}{B}\frac{\partial B_\phi}{\partial z}\ .
\end{equation}
We see that the direction of magnetic flux transport is mainly determined by the
sign of $\eta_H$ (which is positive in general), and the sign of the vertical gradient
of the toroidal field.

In disks, the sign of toroidal magnetic field is generally opposite to the sign of radial field
because of Keplerian shear. For a large-scale poloidal field that drives an MHD disk
wind, the general poloidal field configuration is illustrated in Figure \ref{fig:cartoon}.
As the poloidal field bends away from the protostar, the system tends to develop
oppositely directed toroidal fields above and below the midplane. Therefore, a radial
current is generated in the midplane region, and magnetic flux near the midplane is
transported to the direction that is opposite to this radial current.

Most interestingly, the direction of flux transport is opposite for poloidal fields with
different polarities. Transport is directed inward when poloidal field is aligned with disk
rotation, while it is directed outward for the anti-aligned case. We can also estimate the
rate of magnetic flux transport. Assuming toroidal field varies on scales of disk scale
height $H=c_s/\Omega$, where $c_s$ is gas sound speed, then we find magnetic flux
travels at the radial Hall-drift speed
\begin{equation}\label{eq:hallrate}
|v_{H,R}|\sim \frac{\eta_H}{H}\frac{B_\phi}{B}\approx v_A\frac{l_H}{H}\ ,
\end{equation}
where we have taken $B_\phi\sim B$.
Therefore, if the Hall length $l_H$ is comparable to $H$,
magnetic flux is transported at about the Alfv\'en speed near disk midplane. This
represents a very significant contribution that was largely overlooked in previous
studies.

In comparison, diffusive transport of magnetic flux due to resistivity and AD generally
points outward, at the rate of $v_{O,R}\sim \eta_O/H$ or $v_{AD,R}\sim\eta_A/H$
around the disk midplane (i.e., \citealp{Lubow_etal94a}, and from Equation (\ref{eq:vAD})).
Overall, the three non-ideal MHD effects affect magnetic flux transport in different ways,
while the rate of the transport scale with the respective diffusivities in a similar fashion.

While we discussed the physics assuming there were no charged grains, the
expressions (\ref{eq:vH}) and (\ref{eq:vAD}) are general, with magnetic diffusivities
replaced by more complex expressions (\citealp{Wardle07,Bai11a}). Also note
that $\eta_H$ can change sign for sufficiently strong magnetic field in the presence
of charged grains \citep{XuBai16}, and in that case, the direction of magnetic flux
transport would be reversed.

\section[]{Method}\label{sec:setup}

Transport of magnetic flux is intimately connected to the global disk dynamics.
We proceed to perform global simulations of PPDs to study magnetic flux transport
in a more quantitative and self-consistent manner.

\subsection[]{Simulation Setup}\label{ssec:setup}

We use Athena++, a newly developed grid-based higher-order Godunov MHD code
with constrained transport to conserve the divergence-free condition for magnetic
fields (Stone et al., in preparation). It is the successor of the widely used Athena MHD
code \citep{GardinerStone05,GardinerStone08,Stone_etal08}, and is highly optimized
in several aspects. In particular, it employs flexible grid spacings, allowing simulations
to be performed over large dynamical ranges. Geometric source terms in curvilinear
coordinate systems (e.g., cylindrical and spherical-polar coordinates) are carefully
implemented, which ensures exact angular momentum conservation.

Using Athena++, we solve the standard MHD equations in conservation form
\begin{equation}
\frac{\pa\rho}{\pa t}+\nabla\cdot(\rho{\mb v})=0\ ,
\end{equation}
\begin{equation}
\frac{\pa(\rho{\mb v})}{\pa t}+\nabla\cdot\bigg(\rho{\mb v}{\mb v}
-\frac{{\mb B}{\mb B}}{4\pi}+{\mathsf P}^*\bigg)=-\nabla\Phi\ ,
\end{equation}
\begin{equation}
\frac{\pa E}{\pa t}+\nabla\cdot\bigg[(E+P^*){\mb v}
-\frac{{\mb B}({\mb B}\cdot{\mb v})}{4\pi}\bigg]=-\Lambda\ ,
\end{equation}
where $P$ is gas pressure, $P^*=P+B^2/8\pi$ is total pressure,
$E=P/(\gamma-1)+\rho v^2/2+B^2/8\pi$ is total energy density, $\gamma$
is the adiabatic index, ${\mathsf P}^*\equiv P^*{\mathsf I}$ with
${\mathsf I}$ being the identity tensor, $\Phi=-GM/r$ is the gravitational
potential of the protostar, and $\Lambda$ is the cooling
rate. These equations are coupled with the induction equation (\ref{eq:ind}),
which incorporates non-ideal MHD effects, to evolve the magnetic field.
Also note that in the code, factors of $4\pi$ are absorbed into the definition
of $B$ so that magnetic permeability is $\mu=1$.

With these advantages, we perform 2D global MHD simulations of PPDs in
spherical-polar coordinates ($r-\theta$). The radial grid spans from $r=1$ to
$100$ in code units with logarithmic grid spacing. The $\theta$ grid extends
from the midplane all the way to near the poles (leaving only a $2^\circ$ cone
at each pole), with non-uniform grid spacing where $\Delta\theta$ increases by
a constant factor per grid cell from midplane to pole, with contrasting factor
of four between the midplane and the polar region. This allows us to
properly resolve the disk, and in the mean time accommodate the MHD disk
wind so that the simulation results are not affected by the outer boundary conditions.
Note that in spherical-polar grid, $\theta$ increases from  $0$ to $\pi$
from the north to the south pole. For notational convenience, we define
$\delta\equiv\pi/2-\theta$, namely, the elevation angle about the midplane.

For initial condition, we adopt a self-similar disk density and temperature profiles
in the form of $\rho=\rho_0(r/R_0)^{-\alpha}f(\theta)$, $T=P/\rho=T_0(r/R_0)^{-1}g(\theta)$.
In code units, we set $\rho_0=T_0=R_0=1$. Further, we take $GM=1$ for the gravity of
the central protostar. Once we specify the dimensionless function $g(\theta)$, which
corresponds to the square of local disk aspect ratio $(H/r)^2$, the density profile can
be solved assuming hydrostatic equilibrium. In the $\theta$-direction, it yields
\begin{equation}
\frac{d\ln F}{d\ln\sin\theta}=\frac{GM}{T_0R_0}\frac{1}{g(\theta)}-(\alpha+1)\ ,
\end{equation}
where we have defined $F(\theta)\equiv f(\theta)g(\theta)$. Hydrostatic
equilibrium is possible when the right hand side is positive, which limits the
maximum value of $g(\theta)$.
Force balance in the radial direction determines $v_\phi$:
\begin{equation}
v_\phi^2=\frac{GM-(\alpha+1)T_0R_0g(\theta)}{r}\ .
\end{equation}
Throughout this paper, we choose $\alpha=2$, corresponding to a surface density
profile of $\Sigma\propto r^{-1}$.

In this work, we consider modestly thin disks with midplane aspect ratio
$H_{\rm mid}/r=0.1$ within $\sim2H_{\rm mid}$ about the disk midplane. It
then increases smoothly towards disk surface and reaches $H/r=0.5$ near the
Pole.\footnote{More specifically, the temperature profile is described by three
parameters: $\delta_{\rm mid}$, $\delta_{\rm cor}$ and $\theta_{\rm trans}$
\begin{equation}
\begin{split}
g(\theta)=\bigg\{\delta_{\rm mid}+\bigg[\delta_{\rm cor}-\delta_{\rm mid}
+&(0.5-\delta_{\rm cor})\frac{{\rm Max}(\delta\theta,0)}{\pi/2-\theta_{\rm trans}}\bigg]\\
\cdot&\frac{{\rm tanh}(\delta\theta/\delta_{\rm mid})+1}{2}\bigg\}^2\ ,
\end{split}
\end{equation}
where $\delta_{\rm mid}\equiv H_{\rm mid}/r=0.1$, $\theta_{\rm trans}$ is the transition
angle above/below the midplane around which temperature increases, $\delta\theta$
is the angle between $\theta$ and $\theta_{\rm trans}$ (increasing towards the pole).
This functional form allows the disk aspect ratio to increase from $\delta_{\rm mid}$
to $\delta_{\rm cor}$ via a hyperbolic tangent transition, followed by a linear transition
to 0.5. We take $\delta_{\rm cor}=0.3$ and $\theta_{\rm trans}=0.3$.
}
This allows the gas density to drop much more slowly with $\theta$ and hence
alleviates the timestep constraint in the MHD wind zone. Physically, this is motivated
by the fact that the wind zone to be strongly heated by external UV and X-rays that
becomes significantly hotter than the disk interior (e.g., \citealp{Glassgold_etal04,Walsh_etal10}).
Part of the temperature profile $g(\theta)$ can be found in left panels of Figure
\ref{fig:profiles}.
We adopt an ideal gas equation of state with adiabatic index $\gamma=5/3$,
which appropriate for atomic gas in the wind zone, but the results are insensitive to
the choice of $\gamma$. We seek for simple prescriptions of thermodynamics,
achieved by a simple cooling prescription (the $\Lambda$ term) that relaxes gas
temperature to the initial value at the rate of local Keplerian frequency (based on
spherical radius $r$). This prescription also avoids the development of hydrodynamic
instabilities (e.g., \citealp{Nelson_etal13}).

Poloidal magnetic fields are initialized with vector potential generalized from
\citet{Zanni_etal07}
\begin{equation}
A_\phi(r,\theta)=\frac{2B_{z0}R_0}{3-\alpha}\bigg(\frac{R}{R_0}\bigg)^{-\frac{\alpha-1}{2}}
[1+(m\tan\theta)^{-2}]^{-\frac{5}{8}}\ ,
\end{equation}
where $R\equiv r\sin\theta$, and $m$ is a parameter that specifies the degree that
poloidal fields bend, with $m\rightarrow\infty$ giving a pure vertical field. Poloidal
field is given by ${\mb B}=\nabla\times(A_{\phi}\hat{\phi})$, so that in the midplane,
${\mb B}=B_{z0}\hat{z}(r/R_0)^{-(\alpha+1)/2}$, maintaining constant ratio of gas to
magnetic pressure, defined by plasma $\beta_0$. In practice, we choose $m=0.5$,
and we have tested that the results are insensitive to the choice of $m$.
Fiducially, we choose $\beta_0=10^4$, appropriate for the outer
region of PPDs \citep{Simon_etal13b,Bai15}, but we also consider stronger
and weaker fields in Section \ref{sec:beta}.

\begin{figure*}
    \centering
    \includegraphics[width=180mm]{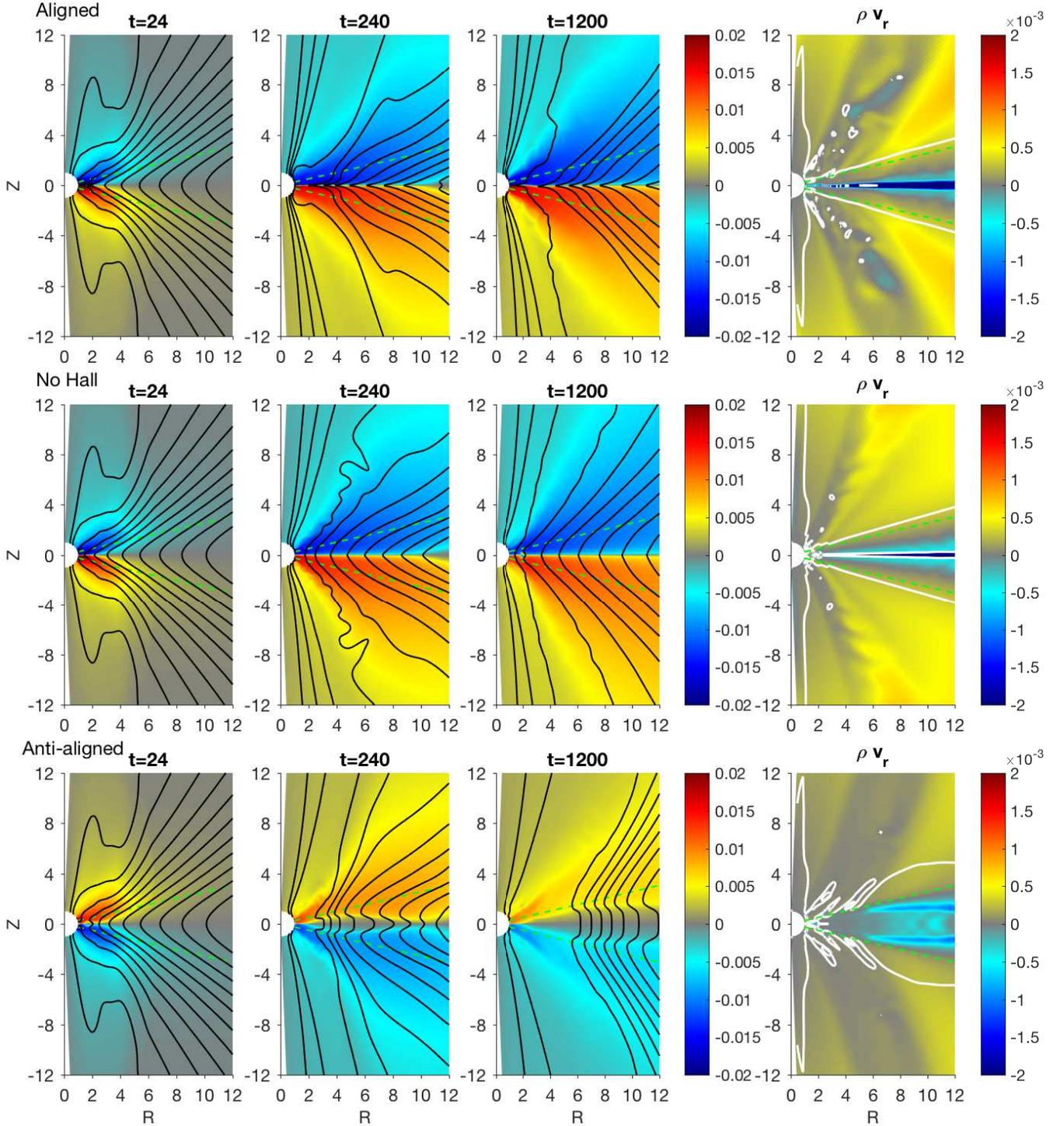}
  \caption{Left three columns: snapshots of magnetic field configuration represented
  by equally-spaced contours (of poloidal magnetic flux) and color (toroidal field $RB_\phi$)
  at $t=24$, $240$ and $1200\Omega_0^{-1}$. Right column: radial mass flux $\rho v_r$
  (rescaled by $r^{-\alpha}R^{-1/2}$), overlaid with the Alfv\'en surface marked by
  white contours. Top and bottom panels are from simulations with poloidal field
  aligned and anti-aligned with disk rotation (Fid$\pm$), while the middle panels are
  from run Fid0 without including the Hall effect. We also include green dashed
  lines which mark an opening angle of $\delta=0.25$ radian above/below disk midplane, and
  it roughly corresponds to the transition from the non-ideal MHD dominated disk zone to
  the disk ``corona" where ideal MHD applies.}\label{fig:Bevolve}
\end{figure*}

We have implemented all three non-ideal MHD terms in Athena++.  In particular, Ohmic
resistivity and ambipolar diffusion are implemented using operator splitting, as in the
original Athena code \citep{BaiStone11} with super-timestepping \citep{Simon_etal13a}.
We have tested the operator-split implementation of the Hall term following \citet{Bai14},
which was shown to be marginally stable in Cartesian coordinates, and found that it
becomes unstable in spherical coordinates. We thus adopt the non-operator-split
implementation following \citet{Lesur_etal14}, where the Hall term is incorporated to the
Harten-Lax-van Leer (HLL) Riemann solver. While the HLL solver is very diffusive,
recent shearing-box simulations using this method \citep{Lesur_etal14,Simon_etal15b}
have yielded results consistent with \citet{Bai14,Bai15}, who used operator-split and the
more accurate HLLD solver \citep{MiyoshiKusano05}.

In this paper, we focus on regions of PPDs where the Hall effect and ambipolar
diffusion are the dominant non-ideal MHD effects, and only include the two terms
in our simulations. It typically corresponds to
regions of intermediate radii ($r\sim5-30$ AU) \citep{Wardle07,Bai11a}.
We set $Am=0.5$ throughout the disk zone (within $\delta\sim\pm2(H_{\rm mid}/r)$),
which is motivated from ionization chemistry calculations \citep{Bai11a,Bai11b}
where $Am$ is found to be of order unity in the outer disks and becomes
smaller towards the inner disk. We take $Am$ to be on the low side which helps
suppress the MRI, or at least to significantly reduce its growth rate so that the
gas remains largely laminar during the simulations.
We further set $l_H=2H_{\rm mid}$ in the inner edge of the disk midplane,
which is taken from estimates in \citet{Bai15}. The value of $l_H$ elsewhere is simply
determined from the relation $\eta_H/\eta_A\propto(B/\rho)^{-1}$, which leads to
$l_H\propto r^{1/2}\sqrt{\rho}$ for constant $Am$. Beyond
$\delta\sim\pm2(H_{\rm mid}/r)$, we smoothly reduce both Hall and AD diffusivities to
zero, mimicking the fact that external far-UV ionization substantially increases the
ionization fraction which brings the gas to the ideal MHD regime
\citep{PerezBeckerChiang11b}.

Our fiducial simulations are performed with $560\times216$ grid cells in $r\times\theta$.
With non-uniform grid spacing, we achieve a resolution of about $16$ cells per
$H_{\rm mid}$ in $\theta$ and $12$ cells per $H_{\rm mid}$ in $r$ around disk midplane.
For the fiducial runs, we also conduct simulations using twice the resolution for
convergence study. As an initial effort, our simulations are 2D instead of 3D, which on the
one hand substantially reduces the computational cost, and moreover, with the MRI
largely suppressed/damped due to strong AD \citep{BaiStone11}, we expect our 2D
simulations to capture the essential aspects of disk dynamics.

The outer radial boundary follows from standard outflow boundary prescriptions, where
hydrodynamic variables are copied from the last grid zone assuming
$\rho\propto r^{-2}$, $v_\phi\propto r^{-1/2}$, with $v_r$ and $v_\theta$ unchanged
except that we set $v_r=0$ in case of inflow.
At the inner radial boundary, hydrodynamic variables are fixed to initial state. We make
this choice because the flow near the polar region is presumably originated from the part
of the disk that is located within the inner radial boundary, whose dynamics is beyond the
reach of the simulation. An outflow-type boundary condition prescription would
violate causality, which can become unstable in the presence of magnetic fields and
further interfere with the wind flow in the main computational domain.
The fixed state boundary condition alleviates the causality issue., and as the gas
flow in our simulations is largely laminar, it also guarantees stability. This allows us
to pursue our study without being affected from the inner boundary.
Magnetic variables in the inner/outer ghost zones are copied from the nearest grid zone
assuming $B_r\propto r^{-2}$ and
$B_\phi\propto r^{-1}$, with $B_\theta$ unchanged. Moreover, we smoothly reduce
the Hall diffusivity to zero within about $H_{\rm mid}$ from the inner radial boundary
to avoid dramatic flux transport due to the Hall-drift.
Reflection boundary conditions are applied in the $\theta-$boundaries.

\begin{table}
\caption{List of Simulation Runs}\label{tab:runlist}
\begin{center}
\begin{tabular}{c|ccccc|ccc}\hline\hline
 Run & Polarity & Resolution & $H_{\rm mid}/r$ & $\beta_0$ \\\hline
Fid+ & + & $560\times216$ & 0.1 & $10^4$ \\
Fid0 & No Hall & $560\times216$ & 0.1 & $10^4$  \\
Fid$-$ & $-$ & $560\times216$   & 0.1 & $10^4$ \\\hline
Fid-hires+ & + & $1104\times432$ & 0.1 & $10^4$ \\
Fid-hires$-$ & $-$ & $1104\times432$ & 0.1 & $10^4$ \\\hline
B3+ & + & $560\times216$ & 0.1 & $10^3$ \\
B30 & No Hall & $560\times216$ & 0.1 & $10^3$ \\
B3$-$ & $-$ & $560\times216$   & 0.1 & $10^3$  \\
B5+ & + & $560\times216$ & 0.1 & $10^5$ \\
B50 & No Hall & $560\times216$ & 0.1 & $10^5$ \\
B5$-$ & $-$ & $560\times216$   & 0.1 & $10^5$ \\\hline
\end{tabular}
\end{center}
\end{table}

\subsection[]{Simulation Runs}\label{ssec:runs}

We list all our simulation runs in Table \ref{tab:runlist}. We will focus on our fiducial runs
labeled ``Fid$\pm$" in the main text, where the $+$/$-$ signs correspond to simulations
with poloidal field aligned/anti-aligned with disk rotation. For comparison, we also conduct
a run ``Fid0", where we turn off the Hall effect. Results from our high-resolution run are
discussed in Appendix \ref{app:hires} to address numerical convergence. We further
discuss the dependence of flux transport rate on the poloidal field strength in Section
\ref{sec:beta}.

We are most interested in the evolution in the inner part of our
simulation domain where the Hall effect is dominant at the midplane. The simulations
are run for about 400 rotations at the innermost disk radius ($2400\Omega_0^{-1}$ where
$\Omega_0=1$ is the Keplerian frequency at the innermost orbit). This is much longer
than the dynamical timescale within $r\lesssim10-15$ so that the flow structure (e.g.,
disk winds and accretion flow) is approximately steady. On top of such flow structure,
magnetic flux evolves on longer timescales. We call such a situation {\it quasi-steady state},
and over the course we can measure magnetic flux evolution.
Moreover, with weak magnetization, the rate of angular momentum transport is relatively
slow, and within the duration of these simulations, the surface density remains largely
unchanged.

\section[]{Overview of Magnetic Flux Evolution}\label{sec:overview}

\begin{figure*}
    \centering
    \includegraphics[width=180mm]{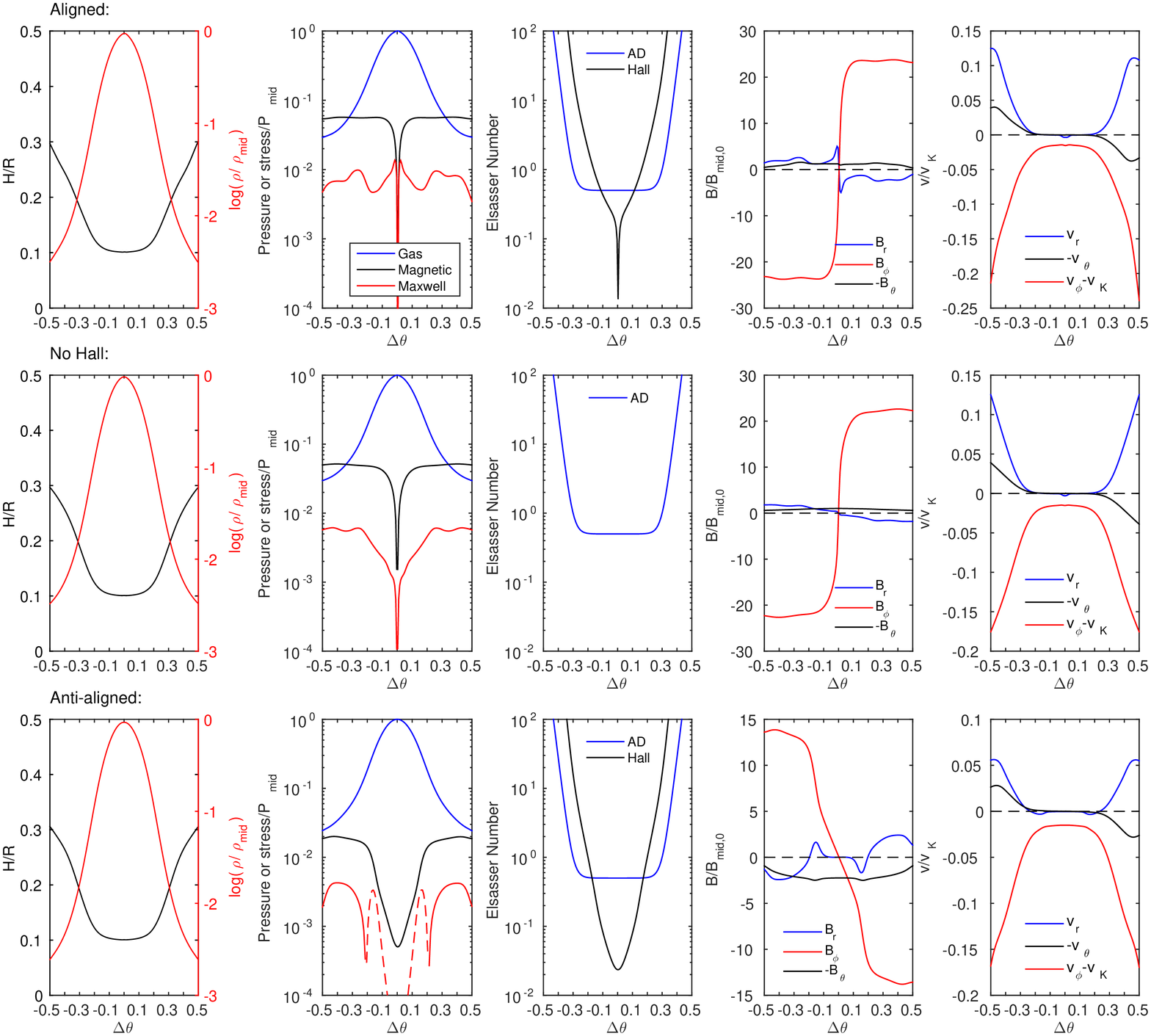}
  \caption{$\theta$-profiles of various hydrodynamic and magnetic variables (as labeled
  in the y-axis and legends) at fixed spherical radius $r=8$, measured at time
  $t=1200\Omega_0^{-1}$. Top and bottom panels correspond to results from simulations
  with aligned and anti-aligned poloidal fields, while the middle panels are from run
  Fid0 with the Hall term turned off. In these plots, gas density is
  normalized to midplane gas density, pressure (gas and magnetic) and the Maxwell
  stress ($-B_RB_\phi/4\pi$) are normalized to midplane pressure $P_{\rm mid}$,
  magnetic field strengths are normalized to initial midplane field strength $B_{{\rm mid},0}$,
  corresponding to plasma $\beta_0=10^4$, and velocities are normalized to Keplerian
  velocity $v_K$. The insets on the rightmost panels zoom in the accretion velocity $v_r$
  in the disk zone.}\label{fig:profiles}
\end{figure*}

In Figure \ref{fig:Bevolve}, we show snapshots of magnetic field configurations for all three
fiducial runs as the systems evolve.
The first snapshot ($t=24\Omega_0^{-1}$) is close to the initial state, where initial field
configuration is best seen at larger radii. During initial evolution, the outward-bent poloidal
field generates oppositely-directed toroidal field above/below midplane due to radial shear,
as discussed in Section \ref{sec:physics}. In the mean time, some of the magnetic flux
that penetrates into the inner radial boundary moves into the polar region and becomes
more vertical, launching a collimated jet. While the jet is irrelevant to our study (and it is
artificial and is not under direct control), it helps stabilize the polar region.

\subsection[]{The Aligned Case}\label{ssec:hallplus}

The early evolution follows exactly from the expectations discussed in Section
\ref{sec:physics}. In the aligned case, the Hall effect efficiently transports magnetic
flux inward at the midplane, leading to some flux accumulation at the inner radial
boundary. Beyond the midplane, however, the toroidal field gradient reverses, and
hence magnetic flux is transported outward. The above two processes stretch the
poloidal field into a highly radially-elongated configuration, as seen in the second
snapshot. As radial field grows, shear produces stronger toroidal field,
which in turn leads to faster flux transport, and further growth of the radial field.
In fact, the runaway process described here is a global manifestation of the
Hall-shear instability \citep{Kunz08}, previously discussed in local shearing-box
simulations \citep{Lesur_etal14,Bai14}. Saturation of this instability is owing to
additional dissipation, and here AD in our simulations, which we
will discuss in more detail in Section \ref{ssec:aligned}.

Upon saturation, the
system achieves a quasi-steady state, with magnetic field bending sharply
across the disk midplane. This configuration allows dissipative processes (here
AD), which generally tend to straighten field lines, to effectively transport
magnetic flux outward in the midplane region. It is the competition between
inward transport by the Hall effect and outward transport by AD that determine
the overall direction of the transport at the midplane: looking at the third snapshot,
the direction of transport is pointing outward.

\subsection[]{The Anti-aligned Case}

In the anti-aligned case, the opposite occurs. We see that first, outward transport
of magnetic flux takes place at disk midplane as a result of shear-produced toroidal
field gradient, as discussed in Section \ref{sec:physics}. The Hall drift pushes the
poloidal field into an unusual configuration which first bends radially inward and then
outward. Later on, poloidal field lines around the midplane straighten. This was
discussed in \citet{Bai14}, where the situation is exactly the opposite to the aligned
case: horizontal components of the field are reduced towards zero instead of
undergoing runaway amplification.

Upon achieving a quasi-steady state, poloidal
field lines near the midplane region are largely vertical, and the toroidal field
gradient across the midplane is greatly reduced but non-zero. With this field
configuration, magnetic flux is transported outward almost entirely due to the
Hall effect at the midplane, with negligible contribution from AD. Towards disk
surface, on the one hand, the Hall effect weakens compared with AD because
of the density drop, and on the other hand, the concavely shaped field configuration
is more favorable for AD to transport magnetic flux outward. Overall, we find that
magnetic flux is systematically transported outward due to the Hall effect at the
midplane and AD at the surface.

\subsection[]{The Hall-free Case}

Without including the Hall term in run Fid0, we see that besides the production of
toroidal field from radial field, there is no significant change of magnetic field
configuration from the initial condition. The system reaches a quasi-steady state
relatively quickly, with toroidal field generation compensated by ambipolar
dissipation in the disk region and outward advection in the disk wind. Despite
the significant difference in field configuration, we find that magnetic
flux is slowly transported outward at a rate comparable to the Fid$\pm$ runs.

\section[]{Gas Dynamics}\label{sec:gasdyn}

In preparation for analyzing the mechanism of magnetic flux transport in more
detail, we discuss the gas dynamics in quasi-steady state in this section.

\subsection[]{Quasi-steady State Gas Vertical Profiles}

In Figure \ref{fig:profiles}, we show the $\theta$-profiles of various diagnostic quantities
of interest measured at spherical radius $r=8$ and time $t=1200\Omega_0^{-1}$
(corresponding to the 3rd snapshot in Figure \ref{fig:Bevolve}). The first (left) column of
panels show the density and temperature (expressed in $H/R$) profiles. The latter is well
preserved from the initial profile due to our cooling prescription. The density
profiles in the bulk disk are largely hydrostatic. Beyond $\delta\sim\pm3H_{\rm mid}/R$,
the density profiles in the aligned and Hall-free cases deviate from the anti-aligned
case. This is related to the fact that magnetic pressure exceeds gas pressure
beyond about $\delta\sim\pm3H_{\rm mid}/R$ in the aligned and Hall-free cases, as
shown in the second column of panels.

The strength of the Hall and ambipolar diffusion terms are measured by their respective
Elsasser numbers, shown in the third panels from left. The AD Elsasser number $Am$ is
independent of field strength, and hence the profiles of $Am$ are the same for all cases.
As we specified in Section \ref{sec:setup}, $Am$ is fixed to $0.5$ in the main disk body
and it increases towards infinity starting from $2H_{\rm mid}$ above/below midplane.
The Hall Elsasser number depends on field strength, and compared with AD, the Hall
term is more important in weaker field. Overall, in both aligned and anti-aligned cases, the
Hall term dominates within about $1-2H_{\rm mid}$ about midplane.

The forth and fifth panels show the profiles for the three components of magnetic fields
and velocities. They are useful for interpreting angular momentum transport in the next
subsection. Also note that the azimuthal gas velocity is always sub-Keplerian due to
pressure support.

\subsection[]{Transport of Angular Momentum and Mass}\label{ssec:amt}

In all cases, MHD disk winds are launched. The MHD winds extract disk angular momentum
vertically via a wind stress $T_{z\phi}\equiv-B_zB_\phi/4\pi$ exerted at the disk surface
(wind base), leading to an accretion rate $\dot{M}\approx(8\pi R/\Omega)T_{z\phi}$.
Approximately, we specify $\delta_b=\pm0.25$ to be the location of the wind
base\footnote{Conventionally, the wind base is set to be located at
where azimuthal velocity $v_\phi$ to be Keplerian \citep{WardleKoenigl93}. This criterion
no longer holds in our simulations and almost all disk regions are sub-Keplerian. This is
largely due to weak magnetization, as well as the relatively warm disk temperature, 
and hence azimuthal velocity generally falls off long the field lines as quickly as Keplerian
rotation\citep{Bai_etal16}.},
which roughly corresponds to the location where $Am\sim1$ and the gas transitions
from being dominated by non-ideal MHD to satisfying ideal MHD conditions (which
permits efficient wind launching).

We see from Figure \ref{fig:profiles} that in the aligned case, with $B_R$ and $B_\phi$
amplified via the Hall shear instability, we measure $T_{z\phi}\approx8.5\times10^{-3}P_{\rm mid}$
at $r=8$, where $P_{\rm mid}$ is midplane gas pressure. Without the Hall effect,
$B_\phi$ is amplified to similar strength but not $B_R$, which remains very small. We
measure $T_{z\phi}\approx5.8\times10^{-3}P_{\rm mid}$, where it is smaller mainly
because of smaller $B_z$ in this run as a result of magnetic flux evolution.
In the anti-aligned case, while horizontal field in the midplane is reduced, $B_\phi$ grows
steadily towards the surface, and we measure $T_{z\phi}\approx6.3\times10^{-3}P_{\rm mid}$
at the wind base, which is only slightly smaller than the aligned case. These results are similar to
those discussed in \citet{Bai14}.

The aligned case also produces a relatively strong Maxwell stress
$T_{R\phi}=-B_RB_\phi/4\pi$ around the disk midplane, leading to non-negligible radial
transport of angular momentum (magnetic braking).
To order of magnitude, the resulting accretion rate is
$\dot{M}\sim(2\pi/\Omega)\int_{-\theta_b}^{\theta_b} T_{R\phi}dz$,
which is about a factor $H/R$ less efficient than wind-driven accretion (for similar stress
levels). Defining $\alpha=\int_{-\theta_b}^{\theta_b}T_{R\phi}dz/\int_{-\theta_b}^{\theta_b}Pdz$,
we find $\alpha\approx0.013$. Comparing $\alpha$ with $T_{z\phi}/P_{\rm mid}$, we see
that $\alpha$ is not sufficiently large to offset the $R/H$ factor to dominate angular momentum
transport, consistent with local studies \citep{Bai14}. In the anti-aligned case, on the other hand,
due to the reduction of horizontal field, radial transport is completely negligible.

\begin{figure*}
    \centering
    \includegraphics[width=180mm]{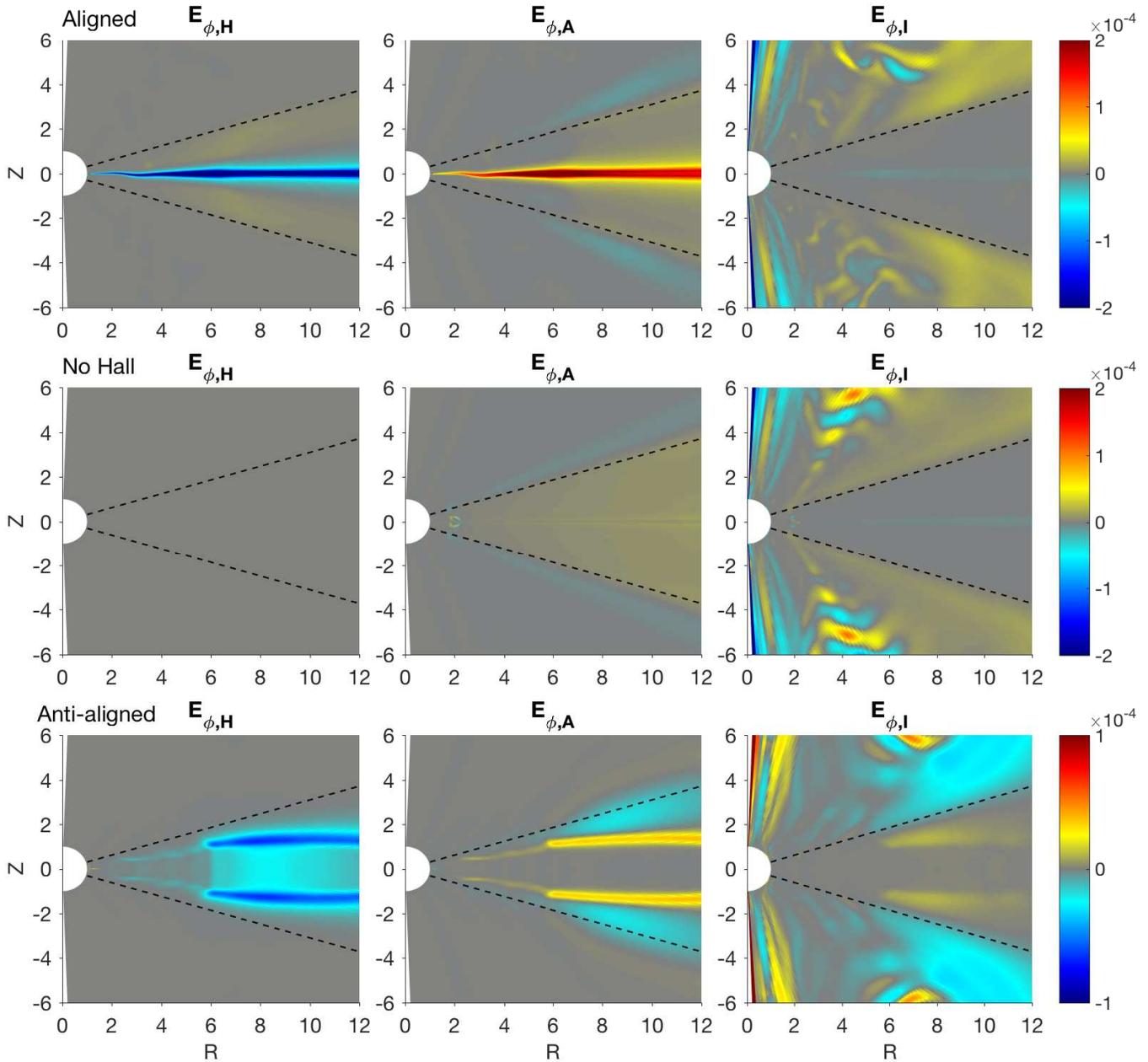}
  \caption{Spatial distribution of the $\phi$-component of the electric field at time
  $t=1200\Omega_0^{-1}$ in our fiducial simulations, which is
  directly related to the rate and direction of poloidal magnetic flux transport. It has been
  rescaled by $R^{-1/2}r^{-(\alpha+1)/2}$. Left, middle
  and right panels show the contribution from the Hall term, AD, and fluid advection in ideal
  MHD. Top and bottom panels correspond to aligned and anti-aligned simulations,
  while the middle panels are from run Fid0 excluding the Hall term.
  Note that in the aligned case, positive (negative) $E_\phi$ means outward
  (inward) transport, while in the anti-aligned case, the opposite holds.}\label{fig:Ephi}
\end{figure*}

The accretion flow associated with the wind can be directly seen in the rightmost panels
of Figures \ref{fig:Bevolve} and \ref{fig:profiles}. Accretion is driven by the torque
associated with the vertical gradient of the wind stress, which is the strongest when
$B_\phi$ varies the fastest. Examining the 4th panels from left in Figure
\ref{fig:profiles}, it becomes clear why the mass flux of the accretion flow is mostly concentrated in
the midplane in the aligned and Hall-free cases, whereas it is more uniformly distributed with
modest concentration around $z=2H_{\rm mid}$ in the anti-aligned case. The total mass accretion
rates in all three cases, on the other hand, are comparable because of their similar
$T_{z\phi}$ values.

The global distribution of mass flux is best viewed from the rightmost panels of Figure
\ref{fig:Bevolve}. The efficiency of wind-driven accretion is characterized by the ratio of
mass loss rate $\dot{M}_{\rm wind}$ to wind-driven accretion rate $\dot{M}_{\rm acc}$.
It is closely related to the location of the
Alfv\'en surface, at which poloidal flow velocity is equal to the poloidal Alfv\'en velocity
$v_{Ap}=B_p/\sqrt{4\pi\rho}$. For wind launched from radius $R_0$, following the field
line, one can define the Alfv\'en radius $R_A$ to be the cylindrical radius at the Alfv\'en
surface. Assuming ideal MHD and axisymmetry in steady state, we have (e.g.,
\citealp{Spruit96,Bai_etal16})
\begin{equation}
\frac{d\dot{M}_{\rm wind}/d\ln R}{\dot{M}_{\rm acc}}=\frac{1}{2}\frac{1}{R_A^2/R_0^2-1}\ .
\end{equation}
The Alfv\'en surface obtained in all our simulations is low, indicating large fractional
mass loss per unit mass of accretion ($R_A/R_0\sim2$). This is related to our adopted
low level of magnetization ($\beta_0=10^4$), and in this case, the winds are largely
driven by magnetic pressure gradient \citep{Bai_etal16}.

We can also see from Figure \ref{fig:Bevolve} that in simulations with the Hall effect,
the change of magnetic field configuration leads to a segregation of magnetic fluxes
in our simulation domain between the inner boundary region and the main disk body.
This is largely a numerical artifact because we do not cover disk regions within the
inner boundary but still have their magnetic flux contained in the polar region.
At $t=1200\Omega_0^{-1}$, there is a lack of magnetic flux between
$R\sim1-6$, where both accretion and outflow mass fluxes tend to diminish, and the
Alfv\'en surface is no longer well defined. We will not discuss this region any further.
Beyond this region, there are also local concentrations/rarefactions of magnetic flux
as a result of intrinsic flux evolution, leading to stronger/weaker local poloidal field
(better seen in the anti-aligned case). They make the Alfv\'en surface move
 further/retreat, which is consistent with theoretical expectations \citep{Bai_etal16}.

\section[]{Magnetic Flux Transport: Detailed Analysis}\label{sec:Btrans}

\begin{figure*}
    \centering
    \subfigure{
    \includegraphics[height=65mm]{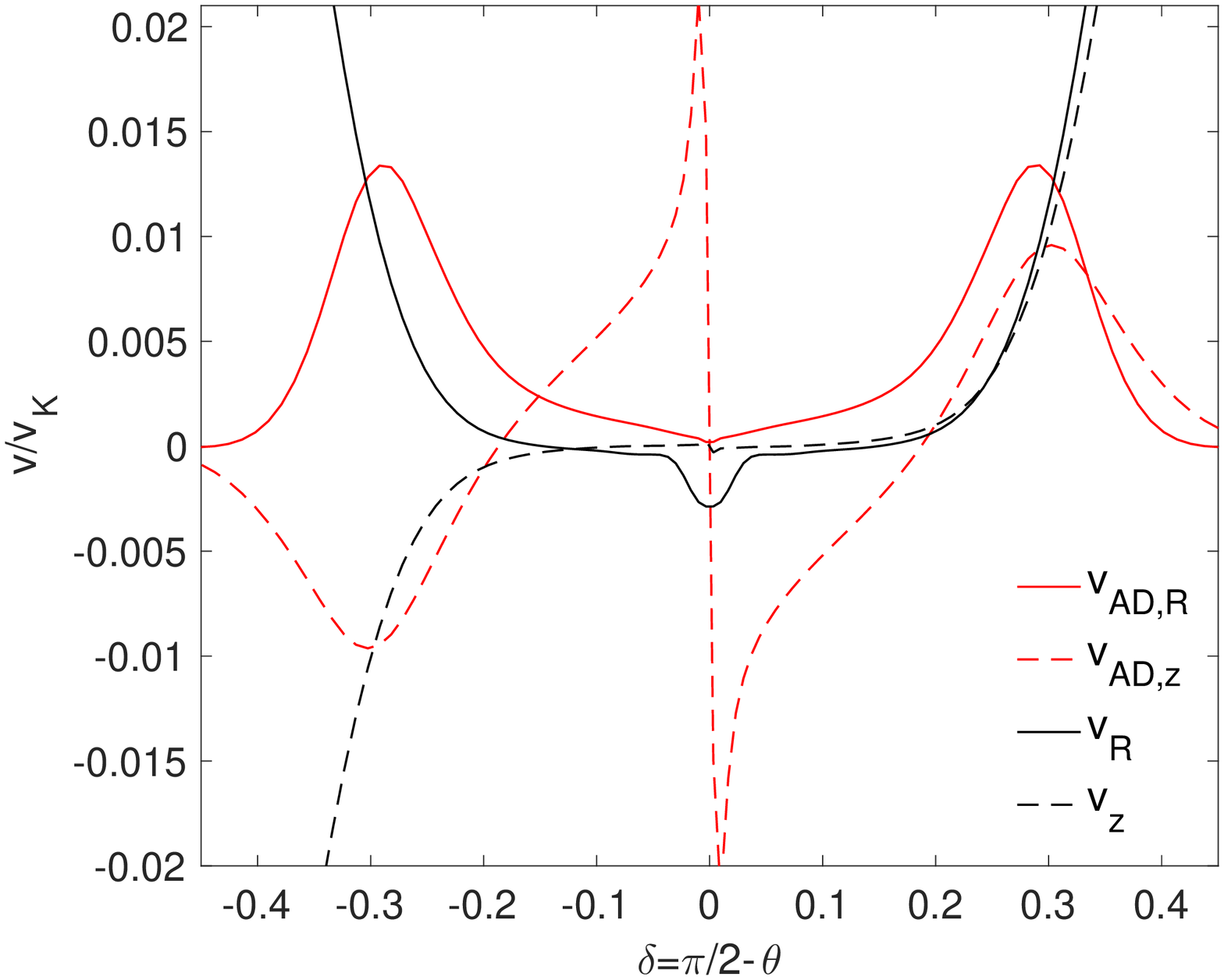}}
    \subfigure{
    \includegraphics[height=65mm]{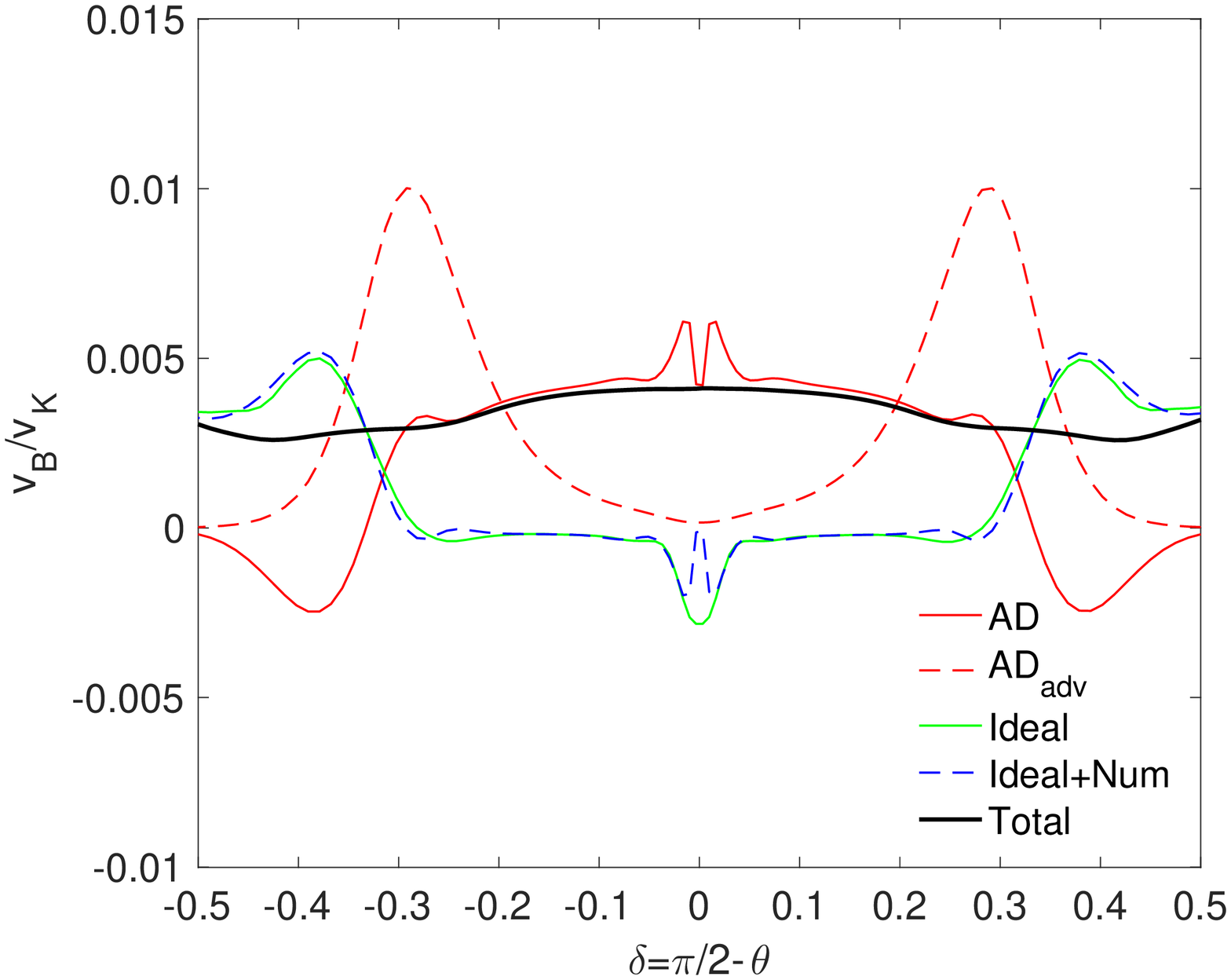}}
  \caption{$\theta$-profiles of various diagnostics that contribute to magnetic flux
  transport measured at spherical radius $r=8$ at time $t=1200\Omega_0^{-1}$ in our
  Fid0 run (no Hall effect).
  Left: decomposition of electron velocity into flow velocity $v$ and ambipolar
  drift velocity $v_{\rm AD}$. Shown are the (cylindrical) radial and vertical components
  of these velocities, as indicated in the legend.
  Right: individual terms in ${\mathcal E}_\phi$ normalized by $v_KB_z$. The
  sum of all contributions is shown in the thick black line. The blue dashed line also include
  contributions from numerical dissipation, and will be discussed in Appendix \ref{app:hires}.
  See text in Section \ref{ssec:nohall} for details.
  }\label{fig:Btrans0}
\end{figure*}

With magnetic flux transport obtained self-consistently with the disk gas dynamics,
in this section, we analyze in
detail the contributions from individual physical effects to the global flux transport.
Our starting point is Equation (\ref{eq:Btrans}). Without explicit
resistivity, flux transport is due to the ${\mb v}_e\times{\mb B}$ term. We decompose
the electron velocity as in
(\ref{eq:ve}), and use (\ref{eq:vH}) and (\ref{eq:vAD}) to compute the Hall drift and
AD drift velocities. The remaining ${\mb v}\times{\mb B}$ term corresponds to fluid
advection as in ideal MHD.
We pick $t=1200\Omega_0^{-1}$ as a fiducial time of evolution, and show the
contributions from these three effects individually in Figure \ref{fig:Ephi}
for our fiducial runs Fid$\pm$ and Fid0. 

\subsection[]{The Hall-free Case}\label{ssec:nohall}

We start from the Hall-free simulation Fid0 as a reference.
Together with Figure \ref{fig:Ephi}, we further show in Figure \ref{fig:Btrans0}
with a more detailed analysis at spherical radius $r=8$. On the left panel,
we show the $\theta-$profiles of radial and vertical components (in cylindrical coordinates)
of ambipolar drift and flow velocities. On the right panel, we show the profile of
\begin{equation}
\frac{v_B}{v_K}\equiv\frac{{\mathcal E}_\phi}{v_KB_z}\ ,\label{eq:vB}
\end{equation}
which is dimensionless and can be considered as the effective velocity of flux transport
$v_B$ normalized by the Keplerian velocity.
Overall, upon reaching a quasi-steady state, magnetic flux is transported
outward ($v_B>0$) at all heights at approximately the same rate. In other words,
the system adjusts/relaxes itself (mainly in its magnetic field configuration) in
such a way that the rate of transport at all heights converges towards a constant value.
We now analyze the contribution from individual physical effects.

\subsubsection[]{Contribution from AD}\label{sssec:AD}

In the bulk disk, we see from Figure \ref{fig:Ephi} and the right panel of Figure
\ref{fig:Btrans0} that outward transport is almost completely due to AD.
For the contribution from AD, we separate out the term $v_{AD,R}B_z$,
denoted as $AD_{\rm adv}$, corresponding to the advection of vertical field due
to radial component of ambipolar drift.
From the left panel of Figure \ref{fig:Btrans0}, we see that $v_{AD,R}$ is positive,
leading to outward flux transport. This transport process is mathematically
(though not physically) analogous to outward transport by Ohmic resistivity discussed
in the literature, and the resulting ${\mathcal E}_\phi$ is dominated by
$\eta_AJ_\phi(B_z^2/B^2)$ (as opposed to $\eta_OJ_\phi$).

However, the rate of transport by this $AD_{\rm adv}$ term is very small near the
midplane. Instead, outward flux transport is dominated by the
$v_{{\rm AD},z}B_R$ term. As we can see from the left panel of Figure \ref{fig:Btrans0},
AD drift near the midplane is dominated by the vertical component, pointing
towards the midplane. This contribution is unique to AD due to its the anisotropic nature.
It results from a strong toroidal field gradient, and by order-of-magnitude, it can be a factor
$\sim(B_\phi/B_z)^2$ stronger than the $AD_{\rm adv}$ term near the midplane.

Physically, this drift motion brings oppositely directed (poloidal and toroidal) magnetic
field into the midplane region.
While the drift velocity $v_{{\rm AD}, z}$
tends to diverge and changes sign at the midplane, this is compensated by the
fact that poloidal field lines becomes more vertical (smaller $B_R$) near the midplane,
and the net rate $v_{{\rm AD},R}B_z$ remains finite at the midplane, as can be seen on
the right panel of Figure \ref{fig:Btrans0}.
Exactly at the midplane where both poloidal and toroidal fields change sign, the vertical
drift velocity vanishes, and outward flux transport there is effectively achieved by
reconnection.\footnote{There are significant contributions from numerical reconnection
in our simulations, but it does not affect the overall rate of flux transport. See Appendix
\ref{app:hires} for further discussion.}

\begin{figure*}
    \centering
    \subfigure{
    \includegraphics[height=65mm]{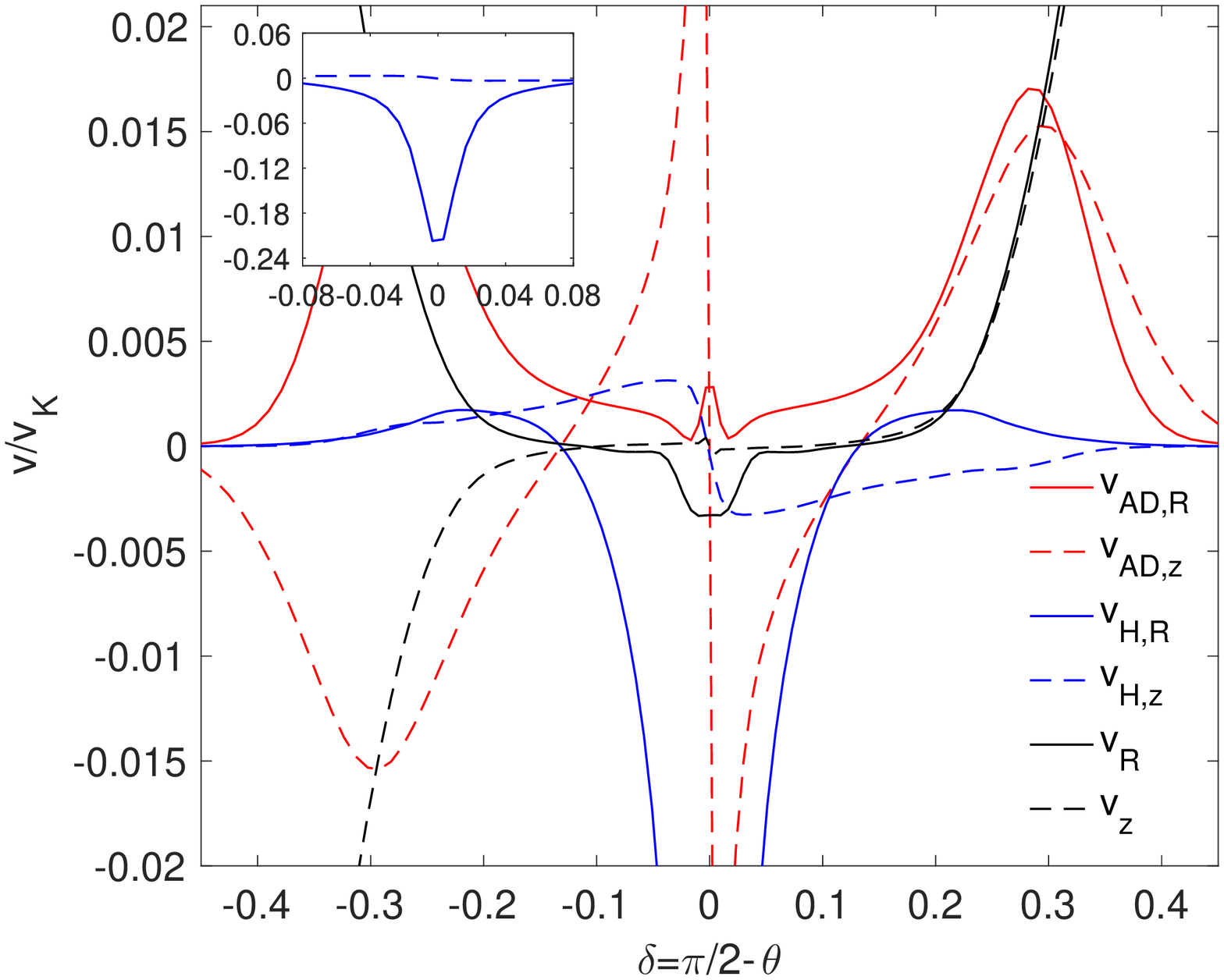}}
    \subfigure{
    \includegraphics[height=65mm]{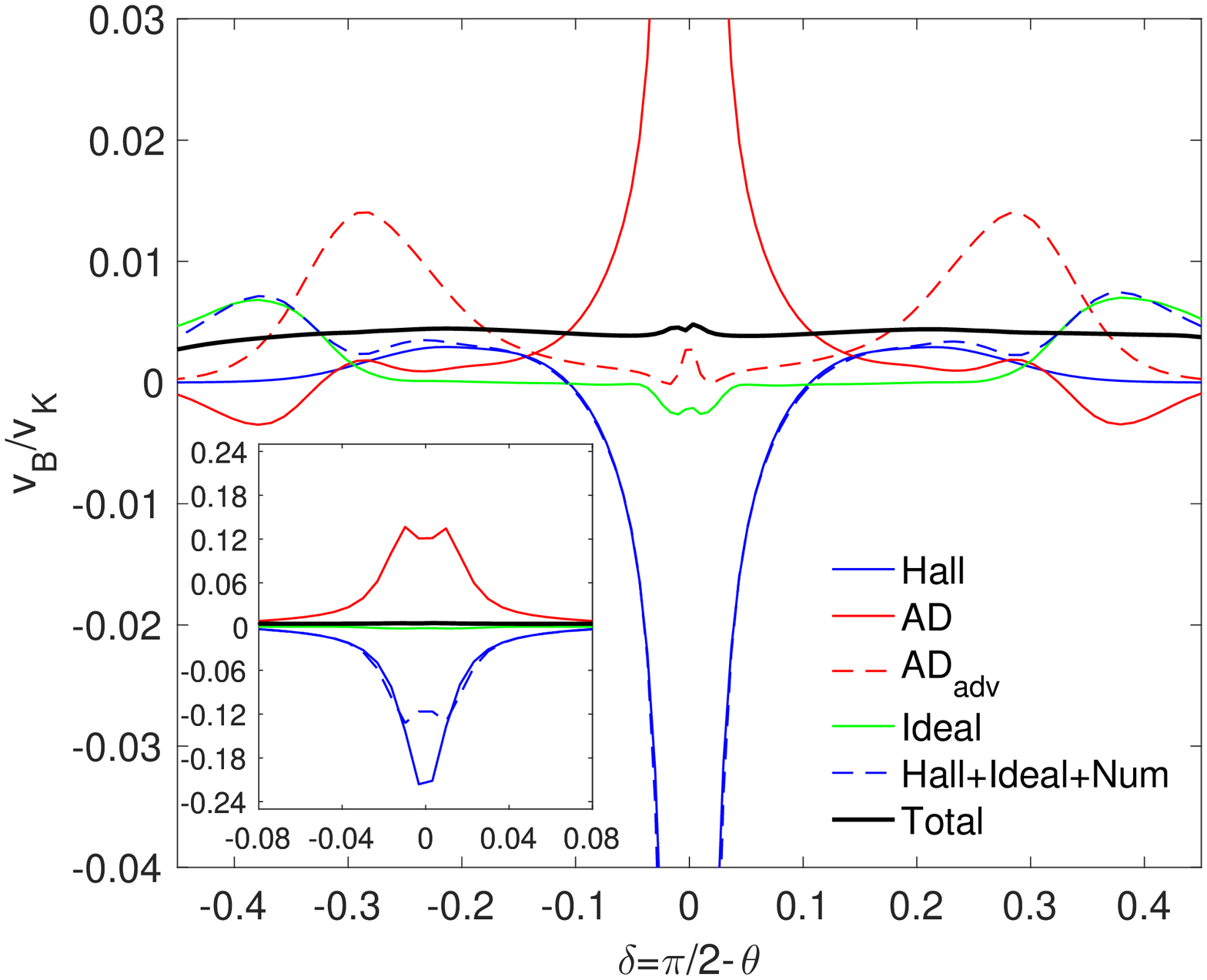}}
  \caption{Same as Figure \ref{fig:Btrans0}, but for run Fid$+$ that includes the Hall effect
  with aligned poloidal field geometry. On the left panel, newly added are lines corresponding to
  radial and vertical Hall-drift velocities $v_{H,R}$ and $v_{H,z}$. On the right panel, newly
  added are the rate of magnetic flux transport due to the Hall drift. 
  The blue dashed line also include contributions from numerical
  dissipation, and will be discussed in Appendix \ref{app:hires}.
  In both panels, an inset is added zooming in the midplane region.}\label{fig:Btransp}
\end{figure*}

\subsubsection[]{Contribution from Fluid Advection}

Fluid advection is associated with bulk gas motion, and in our case, it consists of
the wind-driven accretion flow concentrated at the midplane, and the wind flow itself.
The accretion flow drags magnetic flux inward. However, because the midplane
is the densest part of the disk, accretion velocity is relatively small.
As is shown in the green line on the right panel of Figure \ref{fig:Btrans0}, its
contribution to flux transport is very minor.

In the disk wind zone, we note that in general, fluid advection can
accommodate arbitrary rate of flux transport in any direction. If field lines are all anchored at
fixed radii (no transport), then wind flows travel along poloidal field lines, together with
a series of conservation laws along such field lines (e.g., \citealp{Spruit96}). Non-zero rate
of flux transport can be achieved by having the direction of the wind flow deviate from the
poloidal field direction. This deviation leads to a direct advection of poloidal flux,
corresponding to a non-zero toroidal electric field \citep{Lesur_etal13}. Therefore,
we interpret the outward transport of magnetic flux in the disk wind region mainly as a
response to the transport driven in the midplane region, so as to achieve a quasi-steady
state field configuration. On the other hand, the self-consistent wind dynamics in our
simulation provides the boundary condition for the disk, allowing the field configuration
to adjust itself accordingly.

\subsubsection[]{Overall Rate of Flux Transport}

In quasi-steady state, the rate of flux transport $v_B$ slowly evolves with time, and is
weakly dependent on disk radius (see Appendix \ref{app:global}). For reference, we
quote the value of $v_B\approx4\times10^{-3}v_K$ from our Fid0 run, which
corresponds to the rate measured at $r=8$ and $t=1200$ based on Figure
\ref{fig:Btrans0} averaged within $\delta=\pm0.3$.

We note that this rate is consistent with a simple order-of-magnitude estimate
discussed in Section \ref{sec:physics}:
\begin{equation}\label{eq:vBa}
v_B\sim\eta_A/H\approx\frac{2}{\rm Am}\frac{1}{\beta}\frac{H_{\rm mid}}{R}v_K\ ,
\end{equation}
where the plasma $\beta$ is the ratio of gas to total magnetic pressure. We have
${\rm Am}=0.5$, $H_{\rm mid}/R=0.1$ in run Fid0. The plasma $\beta$ depends
on field strength. Note that while the initial poloidal field has $\beta_0=10^4$, 
in quasi-steady state, toroidal field dominates, and is of the order $\sim10$ times
stronger (see the 4th column of Figure \ref{fig:profiles}) near the midplane.
It gives $\beta\sim10^2$, which together yields $v_B\sim4\times10^{-3}v_K$.

\subsection[]{The Aligned Case}\label{ssec:aligned}

We now consider the Hall simulation Fid$+$ with aligned poloidal field, and
in Figure \ref{fig:Btransp}, we show the contributions to magnetic flux transport
from individual terms at $r=8$ and $t=1200$, with the Hall-drift term (\ref{eq:vH})
included. Again, in this quasi-steady state, magnetic flux is transported outward at
approximately the same rate at all heights.
The role of fluid advection and AD are similar to those discussed in the Hall-free
case. Below, we focus on the Hall-effect mediated flux transport.

\subsubsection[]{Hall-effect Mediated Flux Transport}\label{sssec:indivp}

Recall that the Hall effect is most prominent in the midplane (within $\sim\pm2H_{\rm mid}$),
and its relative importance to AD weakens towards disk surface roughly as $\sim1/\rho$.
Therefore, we focus on the bulk disk region.

\begin{figure*}
    \centering
    \subfigure{
    \includegraphics[height=65mm]{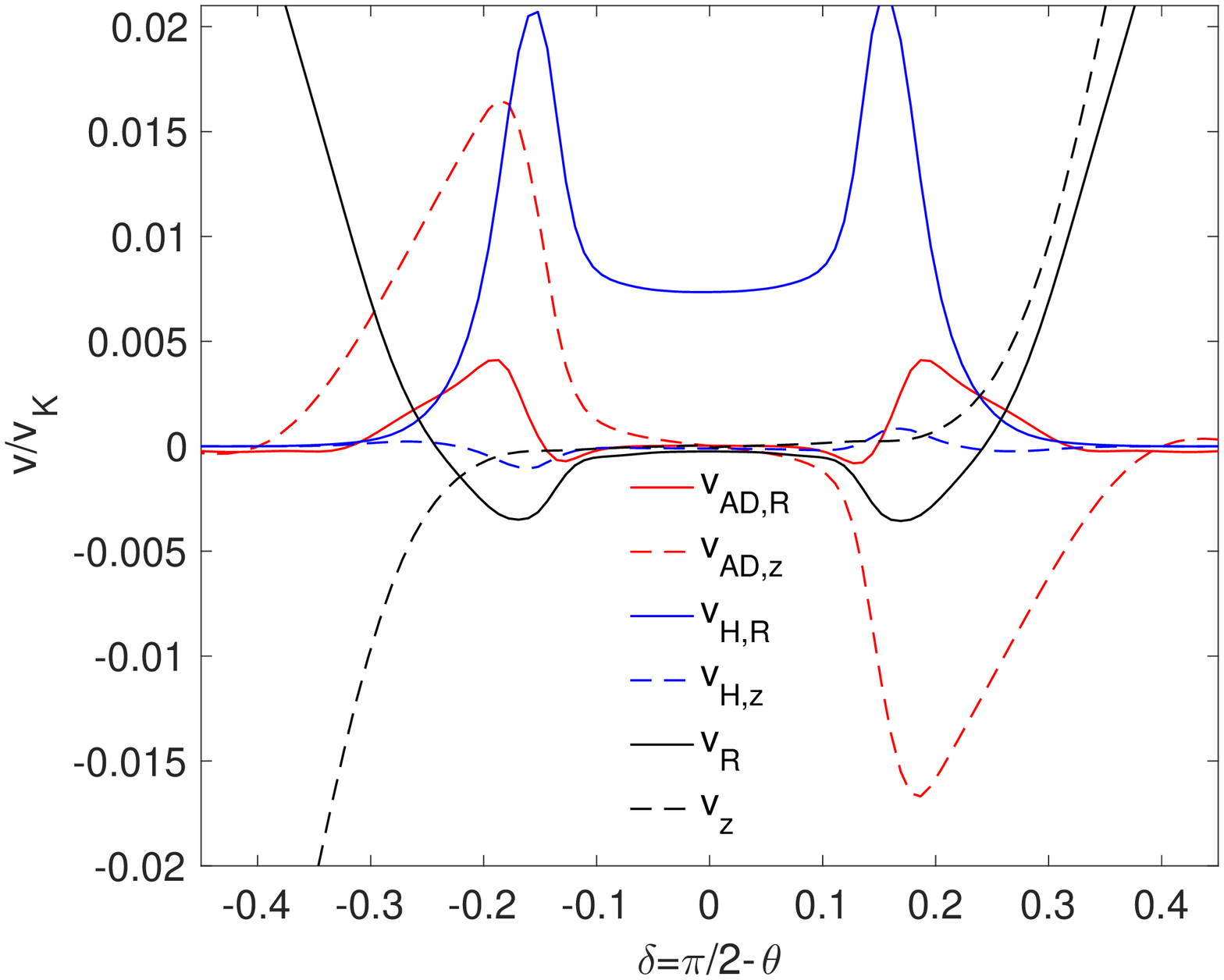}}
    \subfigure{
    \includegraphics[height=65mm]{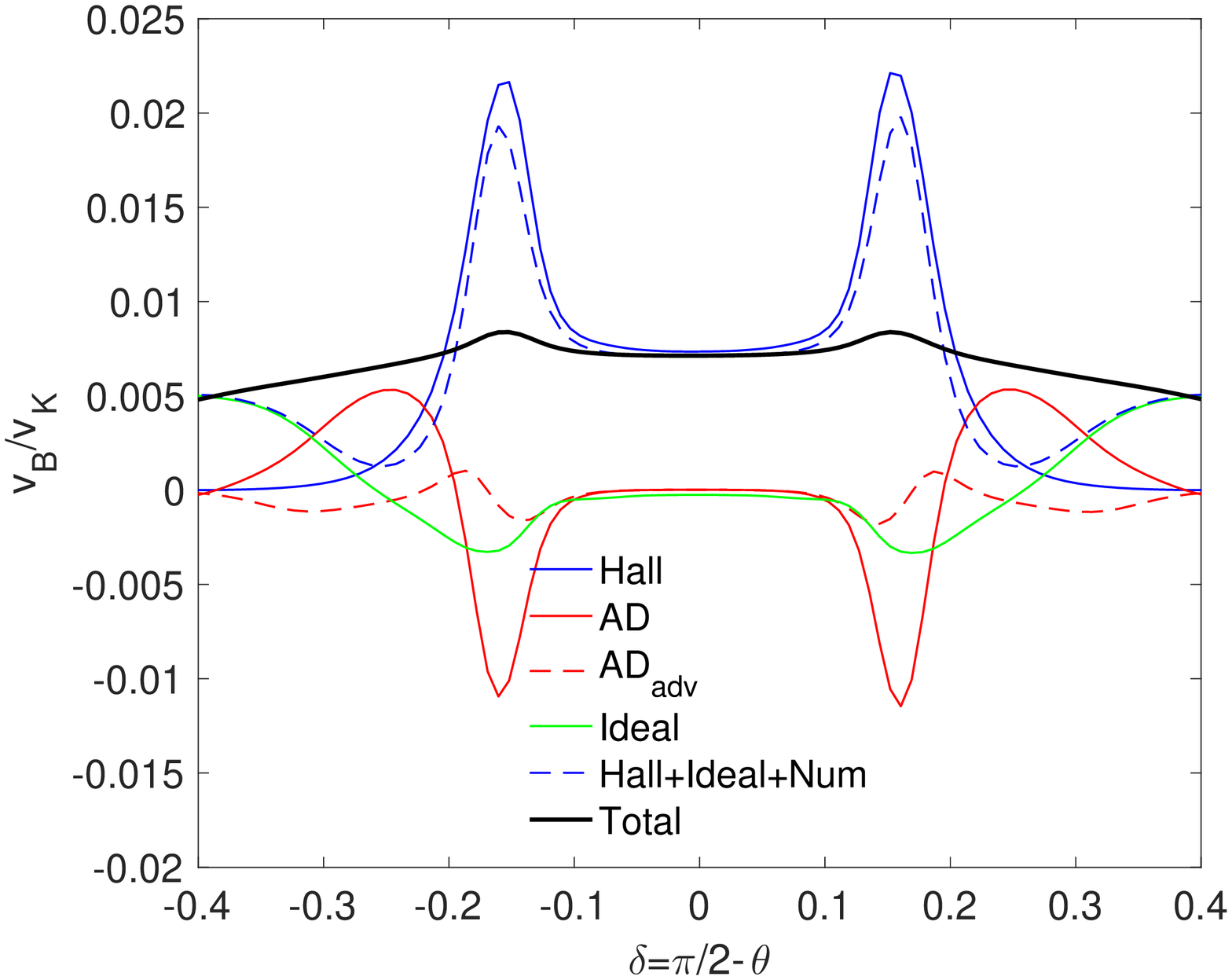}}
  \caption{Same as Figure \ref{fig:Btransp}, but for run Fid$-$ that includes the Hall effect
  with anti-aligned poloidal field geometry.}\label{fig:Btransm}
\end{figure*}

As discussed in Section \ref{sec:overview}, the Hall drift
drags magnetic flux inward in the midplane while pushes the flux outward above and
below. This is confirmed from the top left panel of Figure \ref{fig:Ephi} as well as
Figure \ref{fig:Btransp}. They further show that inward drag is much faster than outward
drift. Inward Hall-drift velocity at the midplane reaches as fast as $\sim20\%$ of the
Keplerian speed! This is largely owing to the strong toroidal field gradient
across the midplane as a result of the Hall shear instability.
Outward transport above/below the midplane due to the Hall effect is only $\sim0.1\%v_K$,
because the (reversed) toroidal field gradient in the upper layer is much smaller,
and the Hall diffusivity is also significantly reduced.
We have also checked that the Hall-mediated transport is almost entirely
due to the $v_{H,R}B_z$ term (radial advection of vertical field). This is because the
toroidal field gradient across the midplane is the strongest, and hence the radial Hall
drift velocity $v_{H,R}$ overwhelms $v_{H,z}$, as seen from the left panel of Figure
\ref{fig:Btransp}.

The highly stretched field configuration across the midplane allows any dissipative
process, in this case AD, to provide outward transport much faster than the Hall-free
case. In fact, it is this outward flux transport that eventually terminates the runaway
growth of the Hall-shear instability, leading to its saturation.
We can see from Figure \ref{fig:Btransp} that outward transport near the
midplane is again dominated by the $v_{AD, z}B_R$ term. Comparing with the
Hall-free case, we see that while $v_{AD,z}$ is only modestly larger, its contribution to
${\mathcal E}_\phi$ from the $v_{AD, z}B_R$ term is much more significant because of
the highly radially-stretched field configuration. Its value slightly exceeds the contribution
from the Hall term ($v_{H,R}B_z$), leading to a net outward transport after
significant cancelation.
In Appendix \ref{app:hires}, we further discuss results from our high-resolution run
Fid-hires$+$ and show that while numerical dissipation also contributes to the outward
flux transport within 2 cells across the midplane, it does not affect the global rate of flux
transport.

\subsubsection[]{Overall Direction and Rate of Flux Transport}

From Figure \ref{fig:Btransp}, we quote the rate of outward flux transport in run Fid$+$
to be $v_B\approx4\times10^{-3}v_K$ measured at $r=8$ and $t=1200$ averaged
within $\delta=\pm0.3$. This is about the same as the Hall-free case, despite the
dramatic influence of the Hall effect.
We now ask, why is magnetic flux eventually transported outward, and at a rate that is
comparable to the Hall-free case?

The direction of flux transport in the midplane region determined by the competition
between the Hall-effect-driven inward transport,
and outward transport due to AD or other dissipative processes (e.g., resistivity). 
Based on our earlier discussion, we expect the rate of outward flux transport to
increase as the radial field becomes more stretched.\footnote{It scales as
$v_{\rm AD}B_R$, where $v_{\rm AD}\sim(dB_{\phi}/dz)B_\phi$. The rate of
Hall-effect-driven inward transport increases as $v_HB_z\sim(dB_{\phi}/dz)B_z$,
which is not as fast.}
In this sense, by adjusting poloidal field configurations, flux transport in both directions
with a wide range of rates can be accommodated.
Therefore, we expect that the global rate of magnetic flux transport is not determined
in the midplane region.

From Figure \ref{fig:Btransp}, we see that at vertical hight $\delta\sim0.1-0.3$
where toroidal field gradient reverses, both the Hall effect and AD lead to outward flux
transport yet no other mechanism can provide any significant inward transport.
We thus conclude that
it is this region that determines the overall rate of flux transport, while other
regions (midplane and the wind zone) adjust their field configuration to achieve the
same rate of transport in response.

To order-of-magnitude, we may estimate the rate of outward transport owing to
the Hall effect using (\ref{eq:hallrate}), but applied to the intermediate layer between
the midplane and the wind zone ($\delta\sim\pm0.2$ in our case):
\begin{equation}\label{eq:hallrate1}
v_B\sim v_A\frac{l_H}{H}=\sqrt{\frac{2}{\beta}}\frac{l_H}{H_{\rm mid}}\frac{H_{\rm mid}}{R}v_K\ .
\end{equation}
At $r=8$ and $\delta=\pm0.2$, we can infer from Figure \ref{fig:profiles} that
$\beta\sim5$, and $l_H\sim0.3H$. We then obtain $v_B\sim2\times10^{-3}v_K$,
which reasonably approximates the measured rate of flux transport.

\subsection[]{The Anti-aligned Case}\label{ssec:antialign}

Finally, we discuss the Hall simulation Fid$-$ with anti-aligned poloidal field, and
show in Figure \ref{fig:Btransm} the contributions to magnetic flux transport
from individual terms at $r=8$ and $t=1200$.
With anti-aligned poloidal field, we find that the mechanism of magnetic flux transport,
as seen from the decomposition shown in the Figures, is qualitatively different
from the Fid0 and Fid$+$ cases.

\subsubsection[]{Contribution from Individual Terms}\label{sssec:indivm}

We start by focusing on the midplane region. As discussed in Section \ref{sec:overview},
the anti-aligned geometry minimizes the vertical gradients of horizontal magnetic field
across the midplane, making poloidal field largely vertical, leaving with only very small
toroidal field gradient \citep{Bai14}. This field configuration reduces the AD drift and fluid
advection velocities to almost zero near the midplane, and they contribute negligibly to
magnetic flux transport. The Hall-drift velocity is also substantially reduced compared to
the case with aligned poloidal field (by a factor of $\sim30$), but owing to its direct
proportionality to $\pa B_\phi/\pa z$, it is the only dominant term present in the midplane.
This fact alone dictates that magnetic flux must be transported outward.

Contribution from the Hall term maximizes at around $\delta\sim\pm0.16$, where
the Hall Elsasser number approaches unity. The reason it is located at this height is
due to a sudden increase of toroidal field gradient (which can be tracked in
Figure \ref{fig:profiles}) and the fact that the Hall
term in the region is still strong enough to provide significant Hall drift.
This enhanced outward transport is partially canceled by inward transport by AD so
that the net rate of transport remains approximately the same as in the midplane.
Transport by AD is again dominated by the $v_{AD,z}B_R$ term. It transports flux
inward in this region because of the unusual poloidal field geometry. While $v_{AD,z}$
still points to the midplane, poloidal fields are bent inward instead of outward, as can
be seen in Figure \ref{fig:Bevolve}. 

Toward the disk surface up to $|\delta|\sim0.25$, the Hall term diminishes and
its contribution rapidly falls off. In the mean time, poloidal field configuration returns normal
and bends outward, where AD leads to outward transport of magnetic flux as usual.
Fluid advection due to wind-driven accretion around $\delta\sim0.2$ accounts for a
higher fraction of the overall rate of transport than the alighed/Hall-free cases, although it
still represents a minor contribution. In the wind zone (beyond $\delta=0.3$), fluid
advection completely takes over to account for the outward flux transport.

\subsubsection[]{Overall Direction and Rate of Flux Transport}

From Figure \ref{fig:Btransm}, we quote the rate of outward flux transport in run Fid$+$
to be $v_B\approx7\times10^{-3}v_K$ measured at $r=8$ and $t=1200$ averaged
within $\delta=\pm0.3$. This is about a factor of 2 faster than the Hall-free and aligned
cases. 
To order-of-magnitude, we may estimate the rate of outward transport from
(\ref{eq:hallrate1}), but using parameters near the midplane region.
We find $l_H/H\sim0.7$, and $\beta\sim10^{2-3}$, which yields
$v_B\sim3-10\times10^{-3}v_K$, consistent with the measured rate of flux transport.

\section[]{Dependence on Poloidal Field Strength}\label{sec:beta}

In this Section, we vary the imposed poloidal field
strength to $\beta_0=10^3$ and $10^5$ with all other parameters fixed (runs B3x
and B5x listed in Table \ref{tab:runlist} where ``x" represents $\pm$ or $0$), and
discuss the results jointly with our fiducial runs with $\beta_0=10^4$.

\begin{figure}
    \centering
    \includegraphics[width=90mm]{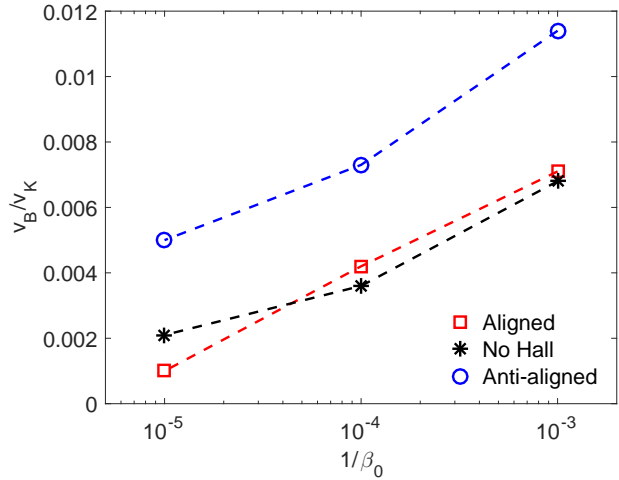}
  \caption{Rate of outward magnetic flux transport $v_B$ as a function of disk
  magnetization (given by the inverse of midplane plasma $\beta_0$). Black asterisks
  correspond to the Hall-free simulations, and red squares and blue circles
  correspond to Hall-simulations with aligned and anti-aligned
  poloidal fields. The rates of transport are measured at $r=8$ and averaged within
  $\delta=\pm0.3$.}\label{fig:parambeta}
\end{figure}

In Figure \ref{fig:parambeta}, we show the measured rate of flux transport $v_B$
from these runs. For consistency, for all runs, we measure the rate of flux transport
at spherical radius $r=8$ at $t=1200$, averaged in between
$\delta=\pm0.3$.
In some of the runs, we find that $v_B$ at fixed radius evolves further after $t=1200$,
which we will discuss in more detail in Appendix \ref{app:global} and argue that the rate
measured at relatively early time ($t=1200$) more appropriately reflects the originally
imposed disk magnetization.

In all cases, we clearly see that $v_B$ increases with stronger magnetization.
Moreover, $v_B$ in the Hall-free case remains to be very similar to the aligned case,
whereas $v_B$ in the anti-aligned case is consistently higher by a factor of $\sim2$.

Qualitatively, the reason behind this trend can be understood from the discussions in
Section \ref{sec:physics}. Given the same magnetic field configuration,
we see that the Hall drift speed $v_{B,H}\propto B$, and the AD drift speed
$v_{B,AD}\propto B^2$ (see also Equations (\ref{eq:vBa}), (\ref{eq:hallrate1})).
In Section \ref{sec:Btrans}, we have identified that the rate of flux transport is
determined by regions at certain heights where the above order-of-magnitude
estimates hold, we thus expect that stronger poloidal field leads to faster transport,
and vice versa.

We also note that the rate of transport mainly scales with the total field strength
$B$ that is dominated by the toroidal component. While total field $B$ increases as
net poloidal field strength increases, the relation between the two is highly nonlinear
as a result of internal disk dynamics. Usually, total $B$ increases more slowly than
net poloidal field. This explains that in Figure \ref{fig:parambeta}, $v_B$ increases
more slowly than the net field strength.

In the conventional theory of magnetic flux transport where flux transport is attributed
to a balance between viscously driven accretion and outward diffusion by resistivity
\citep{Lubow_etal94a}, the rate of transport does not explicitly depend on disk
magnetization. In particular, for outward transport by Ohmic resistivity, the rate of
transport is independent of field strength for a given field configuration. Within the
conventional framework, field strength does indirectly affect the rate of transport by its
feedback on disk dynamics, especially when poloidal field is strong 
(e.g., \citealp{GuiletOgilvie12,GuiletOgilvie13}). Note that the level of magnetization in
all our simulations are relatively weak (since $\beta_0\geq10^3$), yet the dominant role
played by the Hall effect and AD already leads to clearly identifiable dependence of
flux transport rate on disk magnetization.

\section[]{Discussion}\label{sec:discussion}

This work represents the first effort to study magnetic flux transport in PPDs that
incorporates more complex and realistic physics. Our simulations have adopted
idealized prescriptions in disk structure (constant $H/R$), non-ideal MHD diffusivities
(constant $Am$), together with a smooth transition at disk surface. The purpose is
to make the problem reasonably well defined so as to identify important pieces of
physics.

\subsection[]{Relation to the Conventional Theory}

The conventional advection-diffusion framework has been developed further over
the past decade, aiming to address the issue of rapid loss of magnetic flux in thin
accretion disks expected from \citet{Lubow_etal94a}. 
Majority of the works have been constructed under the assumption that the accretion
disk is turbulent as a result of the MRI, with the effect of the MRI turbulence represented
by an effective viscosity and resistivity. Their ratio, called the magnetic Prandtl number,
has been measured to be of order unity in full MRI turbulence
\citep{GuanGammie09,LesurLongaretti09,FromangStone09}, and it leads to the
conventional wisdom that magnetic flux is transported outward for thin accretion disks.
More recently, disk vertical structure and boundary conditions have been realized to
play an important role in the flux transport process. \citet{BisnovatyiLovelace07},
\citet{RothsteinLovelace08} and \citet{BisnovatyiLovelace12}
postulate that the surface of the accretion disks can be largely laminar because it is
magnetically dominated (e.g., \citealp{MillerStone00}) and the MRI is suppressed.
They suggest that inward advection of magnetic flux can be easily achieved in the
non-turbulent, conducting surface layer. \citet{GuiletOgilvie12,GuiletOgilvie13}
presented a family of asymptotic solutions in the thin disk limit assuming
that disk magnetic field is largely vertical. They find that magnetic flux can be
advected inward much faster than the mass flux due to the large radial velocities
at the low-density disk surface, and this effect can substantially alleviate the issue of
magnetic flux loss by outward diffusion. With prescriptions of disk resistivity/viscosity,
the theories have also been applied to PPDs
\citep{Okuzumi_etal14,TakeuchiOkuzumi14,GuiletOgilvie14}, and reasonable
steady-state distributions of magnetic flux have been obtained.

These updated theories may be better applicable to fully MRI turbulent disks (e.g.,
black hole accretion disks), although they remain to be tested against simulations.
However, when applied to PPDs, these theories suffer from several major issues.
First, PPDs are largely laminar with the MRI suppressed or significantly damped
\citep{BaiStone13b,Bai13,Simon_etal13b,Gressel_etal15}. Second, while physical
resistivity is present in PPDs, as considered in \citet{GuiletOgilvie13} and
\citet{Okuzumi_etal14}, it dominates only in very limited regions in PPDs. The Hall effect
and AD, which govern almost the entire range of radii in PPDs, are missing. Third, in the
presence of poloidal field threading the disk, MHD disk wind launching appears inevitable
\citep{SuzukiInutsuka09,Suzuki_etal10,Fromang_etal13,BaiStone13a,BaiStone13b,Lesur_etal13},
which has been ignored in most of the conventional studies (with the exception of Guilet \& Ogilvie
series, who partially incorporated its effect). Besides the wind-driven accretion process,
the wind provides necessary surface boundary conditions that are not easily
incorporated into the existing framework.

Our work advances our understanding of magnetic flux transport in PPDs over
several major aspects. First, we have shown that the Hall effect and AD play distinct
roles in magnetic flux transport that are dramatically different from resistivity.
Because of the anisotropic nature of the Hall effect and AD, magnetic flux transport
depends on the gradient of magnetic fields in all directions, with the most sensitive
being the vertical gradient of the toroidal field. It implies that the process of magnetic
flux transport is strongly coupled with the gas dynamics of the disk itself, leading to
substantial complications.
Second, our simulations have self-consistently incorporated the launching and
propagation of MHD disk winds. On the one hand, it provides realistic ``boundary
conditions" at disk surfaces, and on the other hand, it allows the wind dynamics to
adjust itself to match the rate of flux transport demanded from the disk.

Our results are also in line with some of the updated conventional theories in that
flux transport is mediated by different mechanisms at different heights in the disk.
Therefore, a complete theory of magnetic flux transport in PPDs must properly
take into account the disk vertical structure, and use realistic vertical profiles of
magnetic diffusivities.

\subsection[]{Implications on Disk Formation}\label{ssec:impdf}

Magnetic flux is not only the controlling factor for PPD evolution, it also plays
a fundamental role controlling the process of star and disk formation. Molecular
clouds are known to be relatively strongly magnetized \citep{Crutcher12}, and
the star and disk formation processes naturally inherit some magnetic flux from the
parent cloud. The role of AD has been discussed extensively in the
literature in the context of ``magnetic flux problem" of star formation during core
collapse (see \citealp{McKeeOstriker07} for a thorough review and references therein),
where substantial magnetic flux must be lost. The Hall effect has generally been
considered not to be important during core collapse, as long as there is no significant
rotation to develop strong toroidal magnetic field (e.g., \citealp{KunzMouschovias10}).

Formation of rotationally supported PPDs, on the other hand, unavoidably involve
winding-up of poloidal fields into toroidal fields, leading to significant magnetic
braking. Further loss of magnetic flux is necessary to enable disk formation
(\citealp{MellonLi08}, and see \citealp{Li_etal14} for a review and references therein).
In this context, all non-ideal MHD effects prove to be important
\citep{Li_etal11, Krasnopolsky_etal11,Tomida_etal13}. In particular, it has recently
been found that AD enables significant magnetic flux loss \citep{Tomida_etal15},
and further inclusion of the Hall effect leads to a bimordality on the initial
disk size depending on the polarity of the background magnetic field with respect
to initial angular momentum vector \citep{Tsukamoto_etal15,Wurster_etal16}.

While we have focused on magnetic flux transport in PPDs, the same physics
is applicable in the disk formation process. \citet{Tsukamoto_etal15} and
\citet{Wurster_etal16} have found that a larger disk is formed when field polarity
is anti-aligned, whereas aligned field polarity leads to smaller initial disk size,
although the underlying physical reasons were not addressed.
The fact that both polarities lead to disk formation agrees with our conclusion
that magnetic flux is systematically transported outward. Our finding that the
anti-aligned case loses flux faster than the aligned case implies that less
magnetic flux would be preserved in the former case, and hence weaker
magnetic braking. Therefore, we expect larger/smaller disk to be formed in the
anti-aligned/aligned cases, offering an explanation for the findings of
\citet{Tsukamoto_etal15} and \citet{Wurster_etal16}.

\subsection[]{Global Evolution of Protoplanetary Disks}

Our results serve as a first step towards a better understanding of magnetic flux
transport in PPDs, which largely controls global disk evolution. We discuss below
how our results should be interpreted for this purpose, as well as cautions that
must be exercised.

We first note that the rate of transport obtained in this work is rather high.
For $v_B/v_K=5\times10^{-3}$, magnetic flux depletion timescale would be
on the order of only $\sim30$ local orbits, which amounts to only
$\sim10^3$ years at the distance of 10 AU! However, owing to various caveats
to be summarized in the next subsection, especially the artificial prescriptions
of non-ideal MHD diffusivities, the rate of flux transport measured in this work
is unlikely to be realistic, and readers should not take the values
too seriously. As already emphasized, we aim to clarify the physics before
incorporating more realistic prescriptions, which are left for future works.

The trend that the rate of outward flux transport increases with increasing
net field strength suggests that the strongly magnetized phase of PPDs, if
present, is short lived because of relatively rapid loss of magnetic flux.
The bulk of the disk lifetime is likely associated with relatively weak disk
magnetization. Because stronger magnetic flux leads to higher accretion rate,
our results further imply that the rapid phase of disk evolution with high accretion
rate is brief, and accretion rate decreases with time. As the loss of magnetic flux
slows down over time, we expect the {\it deceleration} of accretion rate to slow
down as well. Although still premature to be incorporated to global disk evolution
models, our results also suggest that the assumption that magnetic flux is conserved
in recent global disk evolution models is unlikely to be valid, while models with
decreasing magnetic flux is more appropriate (see e.g.,
\citealp{Armitage_etal13,Bai16}).

At this point, the speculations above are qualitative. With the current tools available,
it has become feasible to conduct realistic simulations of PPDs that incorporate
more realistic prescriptions, and we expect the physics learned from this
work to be greatly beneficial for future explorations. While we have ignored Ohmic
resistivity in this study, making it more applicable to the outer regions of PPDs
($\gtrsim10$ AU), it is the inner disk ($\lesssim15$ AU) that is expected to be almost
fully laminar \citep{Bai13,Bai14}. A further complication in the inner disk is that
the vertical structure can become asymmetric (e.g,
\citealp{BaiStone13b,Gressel_etal15}). With Ohmic resistivity dominating the
midplane region and suppressing current, the strong current layer typically lies at a
few $\sim H_{\rm mid}$ offset from the midplane. It may be present in only one
side of the midplane, leading to an asymmetric current distribution. When the Hall
effect is turned on, it was found in local shearing-box simulations that maintaining a
physical wind geometry with poloidal field lines bending away from the star can hardly
be achieved \citep{Bai14}. Based on what we have found, such asymmetric current
distribution would lead to asymmetric magnetic flux transport due to the
Hall drift. It is unclear whether a quasi-steady state is possible in such an asymmetric
configuration, which is an intriguing question for future investigations.

\subsection[]{Caveats and Limitations}\label{ssec:cav}

As a first study, our simulations are subject to several caveats and limitations,
some of which are already mentioned earlier in the text and Appendix \ref{app:hires}.
We briefly summarize them below.

A major limitation of the present work is the assumption of axisymmetry. While it
is likely a valid approximation in the inner disk ($r\lesssim10-15$ AU,
\citealp{Bai13,Bai14}), extending it to the outer disk can be problematic because
the MRI can develop in both midplane (weak) and surface (strong), as has been
previously studied in local shearing-box simulations
\citep{PerezBeckerChiang11b,Simon_etal13a,Simon_etal13b,Bai15}. Additional
contribution from
turbulence introduces further complications and uncertainties that need to be
addressed using full 3D simulations.

Another limitation arises from the use of the HLL solver. We have shown in
Appendix \ref{app:hires} that the
rate of flux transport converges with resolution, implying that the system is
able to self-adjust to compensate for the excess numerical dissipation from the
HLL solver. However, it also implies that in reality, the strong current layer
would be much thinner, which raises concerns on its stability. Corrugation of the
strong current layer has already been observed in some of our simulations,
which may eventually destroy the strong current layer.
Similar phenomenon has also been observed in 3D shearing-box simulations of
\citet{Bai15}, where the midplane strong current layer corrugates but without
destroying itself. In general, properly capturing this dynamical behavior would
require full 3D simulations, as well as using less diffusive solvers (at limited
resolution). Therefore, further improvements on the Hall MHD algorithm would
be highly desirable.

For clarity, we have chosen simplified disk models (including thermodynamics)
and prescribed the non-ideal MHD diffusion coefficients. The prescriptions are
motivated from yet do not necessarily reflect realistic disk conditions. In particular,
the non-ideal MHD diffusivities are mainly determined by the disk ionization level,
and the vertical extent where they dominate largely depends on the penetration
depth of external far-UV radiation. None of these processes are included. For
instance, around $5$ AU, the disk would be thinner with $H_{\rm mid}/r\sim0.05$
instead of $0.1$, and non-ideal MHD dominates up to $\sim4H_{\rm mid}$ for
typical FUV penetration depth, instead of $\sim2.5H_{\rm mid}$. Therefore, we
caution on the interpretation of the measured rates of flux transport. Investigations
are underway to use more realistic prescriptions that incorporate ionization
chemistry, far-UV irradiation, flared disk geometry, etc. (see also initial results
from \citealp{Bethune_etal16}).

\section[]{Summary}\label{sec:sum}

In this work, we have studied the transport of poloidal magnetic flux in PPDs,
which has recently been realized to play a crucial role in long-term disk evolution,
and hence many aspects of planet formation. We focus on the regime where the
Hall effect and AD are the dominant non-ideal MHD effects, which is applicable to
a wide range of radii in PPDs. We first demonstrate that the Hall effect in PPDs can
lead to rapid transport of magnetic flux as a result of the Hall-drift, which derives
from a radial current produced from the vertical gradient of toroidal magnetic field.
For typical MHD wind field geometry with toroidal field generated from Keplerian
shear, we expect that in the midplane region, magnetic flux should be transported
inward (outward) for poloidal
field aligned (anti-aligned) with respect to the disk rotation axis.
The rate of transport is proportional to the Hall diffusivity and the vertical gradient of
the toroidal field, and can be on the order of the Alfv\'en velocity for typical disk
conditions.

We then proceed to perform 2D MHD simulations in spherical polar coordinates
in the $r-\theta$ plane using the state-of-the-art Athena++ MHD code. Our simulations
properly resolve the thin disk and in the mean time have the $\theta-$domain
extend to near the polar region to accommodate the launching and propagation
of MHD disk winds. We have implemented and incorporated the Hall effect and AD
in the simulations, with a simple prescription of diffusivities that is roughly applicable
to $\sim5-30$ AU. Our simulation results are valid as long as the disk is largely
laminar, which is likely the case in a wide range of radii for typical PPDs.

Overall, we find that upon reaching quasi-steady states, magnetic flux is systematically
transported outward at approximately the same rate at all heights above/below the
midplane at a given disk radius. The detailed mechanism and the rate of transport,
however, are different for different poloidal field polarities.
The direction and rate of magnetic flux transport is mainly determined by the physics in
the bulk disk, where the transport is mediated by the Hall effect and AD by means of
the Hall drift and ambipolar drift. Wind-driven accretion plays only a minor role in this
process. In the wind zone, flux transport is simply mediated by fluid advection as a
response to the flux transport in the bulk disk, which is achieved by having the wind
velocity vectors deviate from magnetic field lines.

For poloidal field aligned with disk rotation, we find
\begin{itemize}
\item The Hall drift transports magnetic flux inward very rapidly at the midplane,
and outward relatively slowly in the disk upper layer. These processes stretch
the poloidal field into a radially elongated configuration as a global manifestation
of the Hall shear instability.
\item At the midplane, inward transport due to the Hall drift is compensated
by outward transport by AD (dominated by vertical ambipolar drift). Towards disk
upper layer, both the Hall effect and AD contribute to outward transport, which
mainly determines the direction and rate of flux transport.
\end{itemize}

For poloidal field anti-aligned with disk rotation, we find
\begin{itemize}
\item In the midplane region, horizontal field components are suppressed, and
magnetic flux transport is governed by the Hall drift, which points outward.
\item Towards disk upper layer, poloidal field lines bend first radially inward
due to outward flux transport at the midplane, and then outward to launch
the disk wind. The Hall effect (outward), AD, and wind-driven accretion
(inward) all contribute to flux transport.
\end{itemize}

In both cases, and within the parameters explored in this work, we find that outward
transport is inevitable because there are always
regions where only outward transport is possible. These include the disk upper layer
in the aligned case, and the midplane region in the anti-aligned case.
Overall, we find that the anti-aligned case leads to faster outward transport than
the aligned case and the Hall-free case by a factor of $\sim2$. More strongly
magnetized disk leads to faster outward transport.

With our fiducial simulation parameters, we find the net rate of outward transport is
uncomfortably large. We emphasize that the main purpose of this work is to 
demonstrate basic physics, and caution on directly applying the measured rate of
flux transport to PPDs. The resolution to this issue likely lies in the caveats
discussed in Section \ref{ssec:cav}, namely, the need for realistic disk model with
self-consistent ionization-recombination chemistry, and role of the MRI turbulence
which may operate at the surface layer of the outer disk. Our preliminary studies
have found that when realistic diffusivity profile is applied for the inner disk, the
rate of flux transport is significantly slower. In addition, we have also found that
in fully MRI turbulent thin disks, transport of magnetic flux is dominated by
the turbulent disk surface via the ``coronal mechanism", which points radially
inward \citep{Beckwith_etal09}.

Through this work, we point out and clarify the important roles played by the Hall
effect and AD on the transport of magnetic flux in PPDs that have been missing in
conventional theories of magnetic flux transport in accretion disks. The anisotropic
nature of these non-ideal MHD effects makes the flux transport process coupled
with essentially all components of magnetic field gradients (particularly the vertical
gradient of toroidal field), and hence the entire disk gas dynamics. The problem is
intrinsically global and multi-dimensional, requiring MHD disk winds to be well
accommodated, and disk vertical structure to be appropriately resolved.
Future explorations should focus on simulations with more realistic prescriptions
of magnetic diffusivities and disk thermodynamics. We expect the physics learned
from this work to provide key insight as more complexity is built up towards
more realistic studies of PPD.

\acknowledgments

We thank an anonymous referee whose comments helped us improve the
presentation of the paper, and Kengo Tomida for assistance on the
implementation of non-ideal MHD terms. 
X.-N.B is supported by Institute for Theory and Computation, Harvard-Smithsonian
Center for Astrophysics. Computations for this work are performed on the Hydra cluster
managed by the Smithsonian Institution, and on Stampede at the Texas Advanced
Computing Center through XSEDE grant TG-AST140001.

\appendix

\section[]{A: Strong Current Layer at the Midplane and Resolution Study}\label{app:hires}

\begin{figure*}
    \centering
    \subfigure{
    \includegraphics[height=65mm]{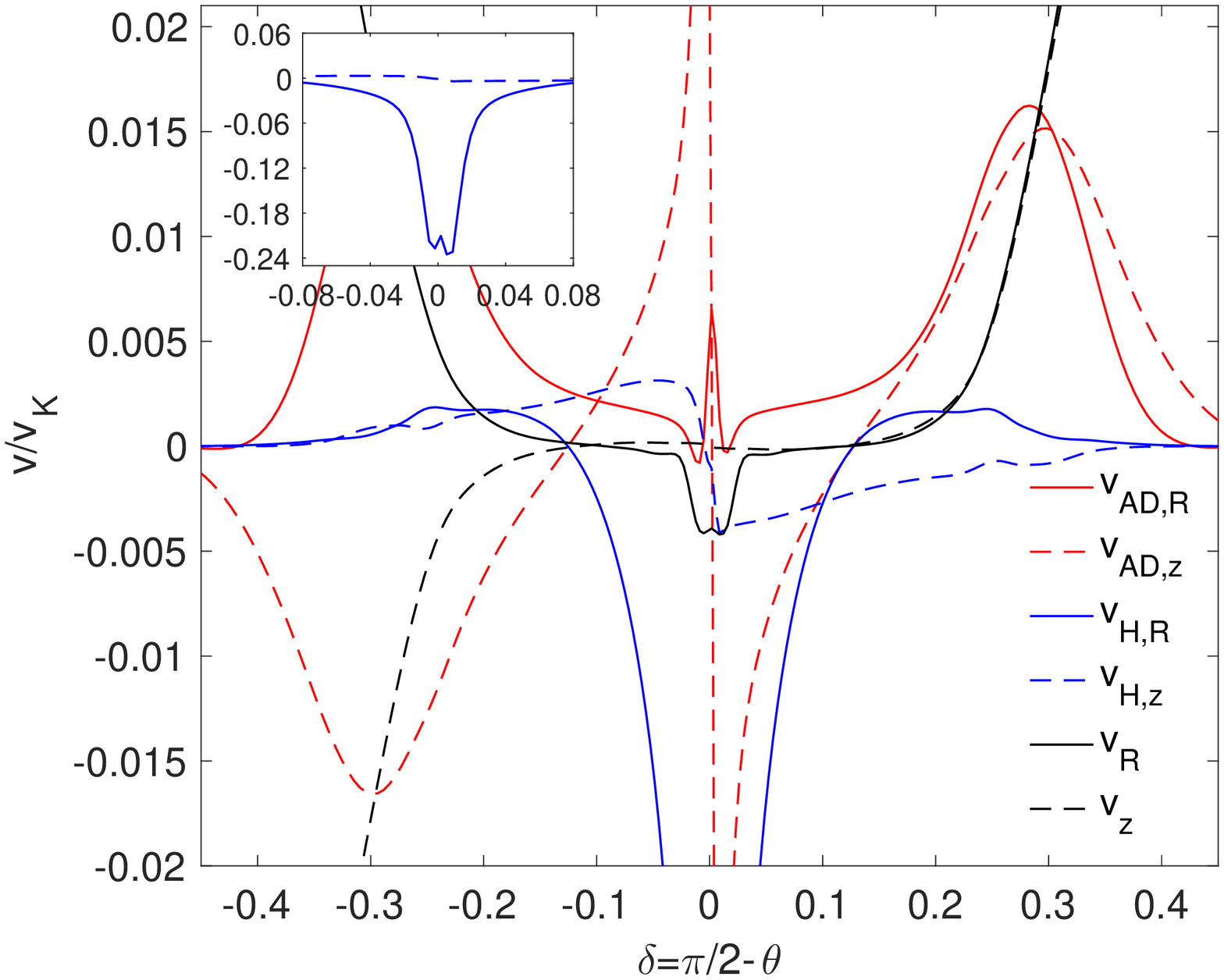}}
    \subfigure{
    \includegraphics[height=65mm]{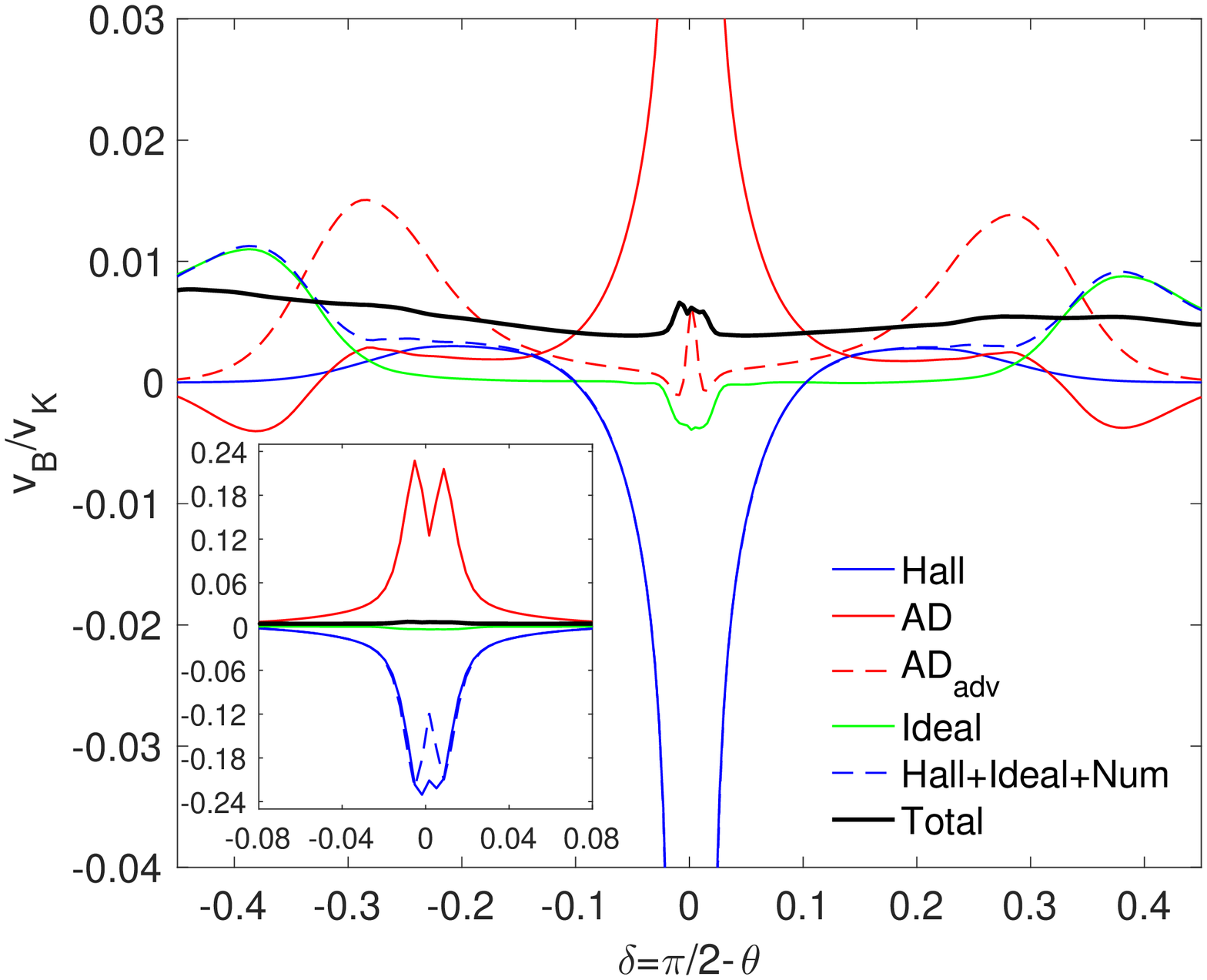}}
  \caption{Same as Figure \ref{fig:Btransp}, but for run Fid-hires$+$ where the resolution is
  doubled.}\label{fig:Btransp_hires}
\end{figure*}

In both Fid0 and Fid$+$ runs, the systems are characterized by a thin,
strong current layer across the midplane. This thin current layer dissipates
shear-generated toroidal magnetic field via reconnection, and leads to
outward transport of poloidal magnetic flux. In our simulations, some of the
reconnection and outward transport are due to numerical dissipation.
In this Appendix, we discuss the physics of the strong current layer, and
show that the presence of numerical dissipation does not affect the
global rate of magnetic flux transport.

Due to the use of the very diffusive HLL solver, numerical dissipation can become
significant in the strong current layer.
In our simulations, we also extract the electric field
${\mathcal E}_\phi$ that is actually used to update the magnetic field via
constrained transport. Their $\theta-$profiles are shown as blue dashed lines on
the right panels of Figures \ref{fig:Btrans0} and \ref{fig:Btransp} for run Fid0 and
Fid$+$, respectively. They contain contributions from the combination of the ideal
MHD and Hall (in run Fid$+$) terms, as well as numerical diffusion. We further
zoom in the profiles near the midplane shown in the insets. By comparing the sum
of blue and green lines with the blue dashed line, we confirm that numerical
dissipation is negligible except in the vicinity of the midplane, which we will focus
on below.

Significant numerical dissipation is localized within $\pm2$ cells
above/below the midplane. In run Fid$+$, inward drag due to the Hall term (blue) is
nearly twice stronger than intrinsic outward transport due to AD, with the rest of the
outward transport owing to numerical dissipation. This dissipation can be effectively
considered as a resistivity. It produces additional ${\mathcal E}_\phi$ from a
toroidal current $J_\phi$, which mainly results from the vertical gradient of $B_R$,
as poloidal field lines bend. Accordingly, the more radially stretched field configuration
in run Fid$+$ leads to much faster numerical transport of magnetic flux compared with
the Fid0 run.

To assess whether numerical dissipation affects the global rate of magnetic flux
transport, we further show in Figure \ref{fig:Btransp_hires} the analysis of the
high-resolution run Fid-hires$+$ where grid resolution is doubled. We see that
contributions from individual terms to magnetic flux transport are almost identical
between runs Fid$+$ and Fid-hires$+$. In particular, the global rates of transport
from the two runs are almost the same (run Fid-hires$+$ yields a value that is
$\sim10\%$ larger at the particular snapshot). This result suggests that the
system is able to adjust its {\it local} magnetic field configuration to adapt to
different levels of numerical dissipation at the midplane without affecting the
{\it global} rate of flux transport, which gives us confidence in our simulation
results.\footnote{We have also repeated simulation Fid0 using the much less
diffusive HLLD solver, and confirm that except for having a sharper current sheet
at the midplane, the rate of magnetic flux transport is also almost the same.}

In reality, in the absence of numerical dissipation, 
the outward transport at the midplane must be mediated by physical dissipation.
Such physical dissipation can be Ohmic resistivity, which is progressively more
important towards the in the inner region of PPDs, and would act in a way
analogous to numerical dissipation. In our case, dissipation is provided by AD.
Note that the $v_{AD,z}B_R$ term discussed in Section \ref{sssec:AD} vanishes
at the midplane, and hence outward transport at the midplane must be due to
the radial advection term $v_{AD, R}B_z$. Although this term appears to be
negligible in run Fid0 (see Figure \ref{fig:Btrans0}), its becomes more noticeable
in run Fid$+$ where the radial field becomes more stretched (see Figure
\ref{fig:Btransp}). With higher resolution, the significance of this term grows
further as the current sheet becomes sharper, as seen from Figure
\ref{fig:Btransp_hires}. In the limit of no numerical dissipation, we thus expect
this radial AD-drift to be able to account for the entire outward transport at the
midplane region. In the mean time, we caution that the development of much
sharper current sheet poses concerns on its stability, which is an important
caveat yet is beyond the scope of the current investigation.

Finally, we briefly comment that in the anti-alignd case, because the magnetic field
profile is much smoother, we find that the electric field returned from the HLL solver
matches very well with the total electric field evaluated separately from fluid advection
and the Hall term, as can be seen from Figure \ref{fig:Btransm}. We have also
conducted a higher-resolution run Fid-hires$-$ and confirm that the results agree very
well with the fiducial run Fid$-$.

\section[]{B: Global Magnetic Flux Evolution}\label{app:global}

In most parts of this paper, we conduct analysis at snapshots where magnetic
flux evolution is in quasi-steady state. In this Appendix, we discuss the time
evolution of magnetic flux during such ``quasi-steady" state.

\begin{figure}
    \centering
    \subfigure{
    \includegraphics[width=58mm]{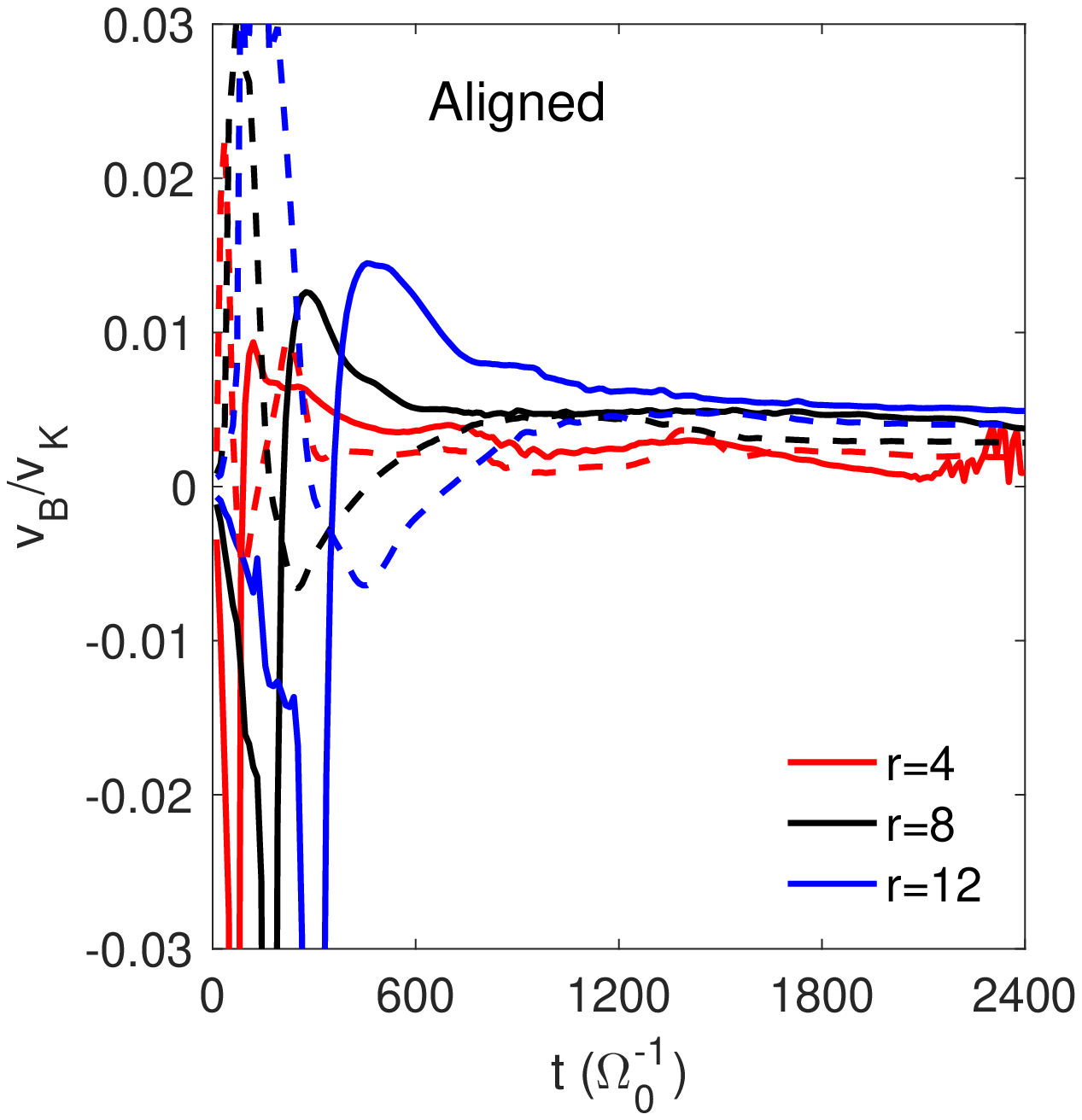}}
    \subfigure{
    \includegraphics[width=58mm]{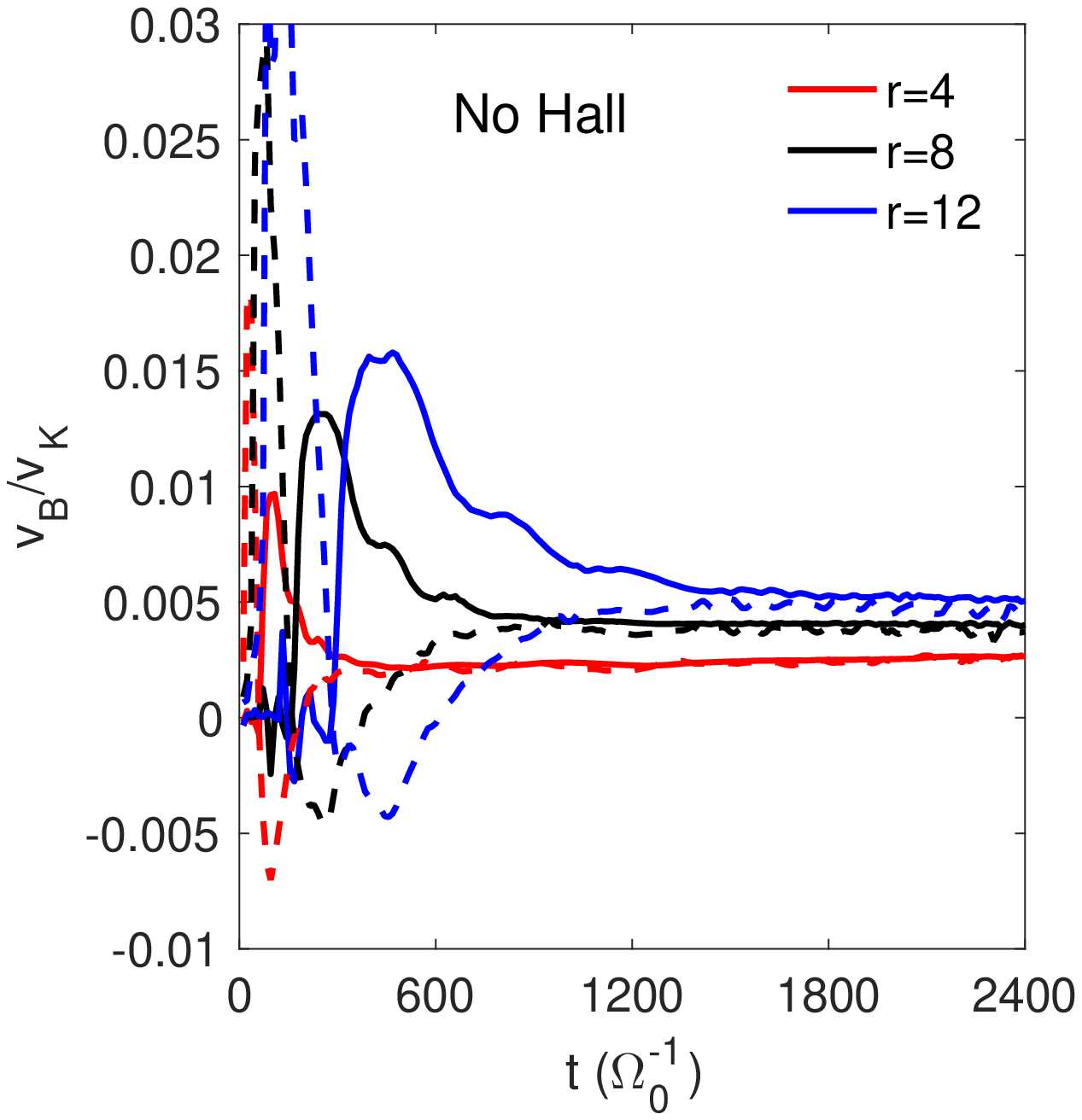}}
    \subfigure{
    \includegraphics[width=58mm]{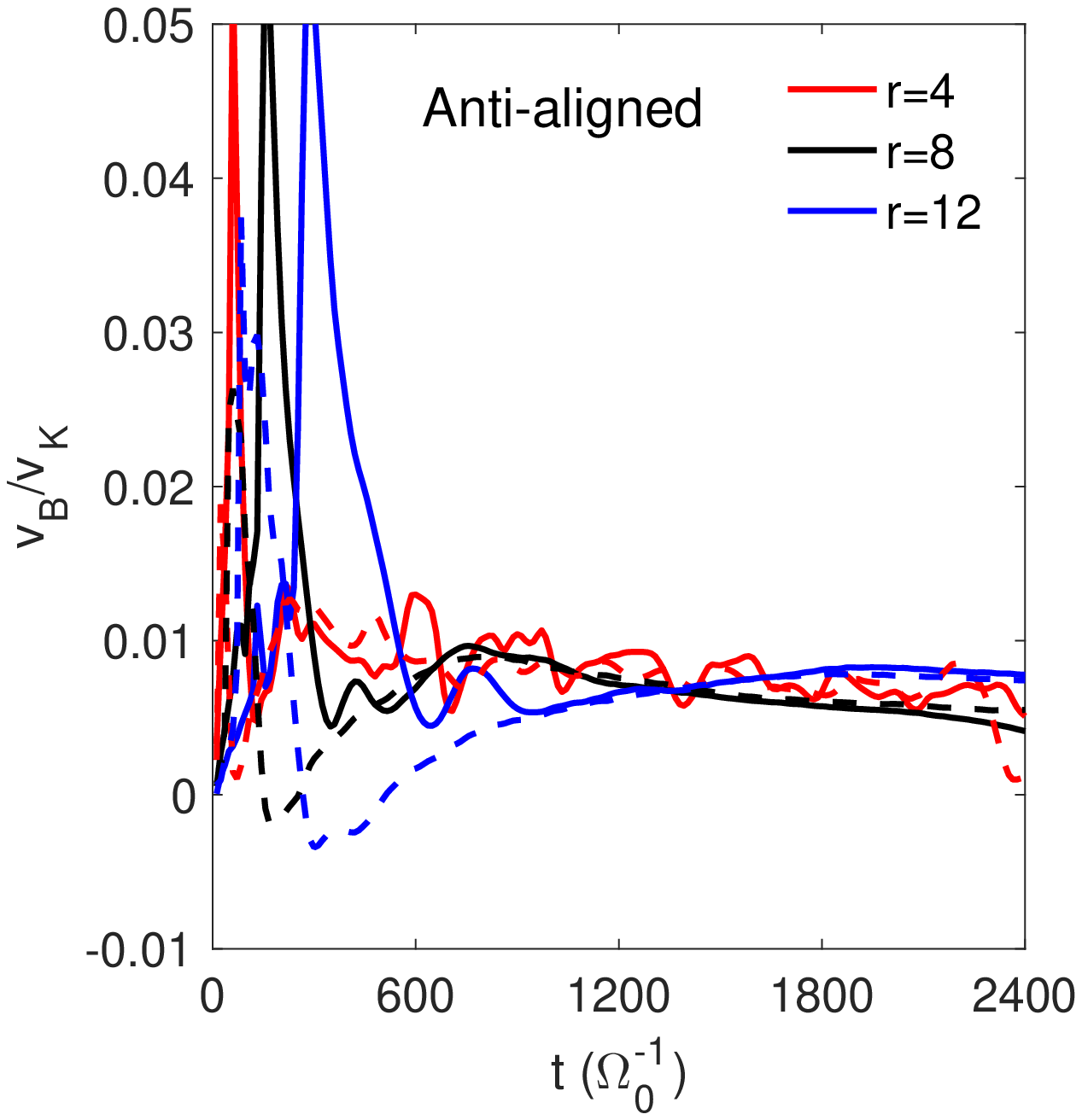}}
  \caption{Time evolution of the rate of magnetic flux transport
  $v_B\equiv{\mathcal E}_\phi/B_z$ in our fiducial run Fid$+$ (left), Fid0 (middle)
  and Fid$-$ (right) at different disk locations as a function of time. Solid lines correspond
  to the midplane regions, and dashed lines correspond to the location $\delta=0.2$
  above/below the midplane (averaged). Line colors represent different disk radii, as
  indicated in the legend.}\label{fig:vBevolve}
\end{figure}

In Figure \ref{fig:vBevolve}, we show the time evolution of $v_B$ defined in
(\ref{eq:vB}), measured at different disk locations including both the disk
midplane and upper layer ($\delta=\pm0.2$ from midplane, averaged), for our
runs Fid$\pm$ and Fid0. 
We see that initially, magnetic flux evolves at different rates between
the midplane and the surface. Later on, after about $\sim4$ local orbital time at each
radius, transport rates at the midplane and upper layer converge and a quasi-steady
state is achieved. Note that the measured rates at midplane and upper layers do not
necessarily match exactly because they are connected to different field lines.

In both the Hall-free case and the aligned case at $r=8$, we see that the rate of flux
transport $v_B$ remains approximately constant over longer-term evolution. This fact
justifies our approach of measuring $v_B$ only at fixed snapshots.
In the anti-aligned case, however, $v_B$ shows some further time evolution.
At our fiducial radius $r=8$, $v_B$ slowly decreases with time, and is
reduced by $\sim40\%$ from time $t=1200$ to the end of simulation at $t=2400$.
The main reason for this reduction is that outward flux transport is the fastest for run
Fid$-$, and towards later time, there is a deficit of flux at small radii. For instance,
we see in Figure \ref{fig:Bevolve} for run Fid$-$ at $t=1200$, magnetic flux is
already substantially depleted in the region around $r=4-6$, and shortly afterwards,
the $r\sim8$ region is also affected. Because of this, the inner disk becomes less
strongly magnetized towards later time, leading to slower flux transport according to
Section \ref{sec:beta}. Similarly, the measured $v_B$ at $r=4$ in the aligned and
anti-aligned cases shows more variability towards later time. This is again due to
flux depletion/segregation at that radius, as can be seen in Figure \ref{fig:Bevolve}.
Therefore, we expect that $v_B$ measured at earlier times more
reliably reflects the true rate of transport at the imposed level of magnetization.

The normalized rate of transport $v_B/v_K$ also shows some radial dependence.
In the Hall-free case, our simulation setup guarantees that the physics is
independent of disk radius, and hence one might expect $v_B/v_K$ should be
independent of $r$. In practice, we see that $v_B/v_K$ is modestly different at
different radii, and is slower at smaller radius. Without this normalization, on the
other hand, we find $v_B$ itself is approximately the same across the range of radii
between $r=4-12$. In the aligned and anti-aligned case where the Hall effect is
included, the measured $v_B$ at $r=8$ and $r=12$ also show some difference
along their evolutionary paths.
Theoretically, magnetic flux transport is intrinsically a global phenomenon, where
the dynamics at different radii can affect each other. Moreover, with magnetic flux
constantly evolving, flux distribution across the disk also varies with time, and
there is no guarantee that the rate of transport has to be constant in time and
radius. As an initial study of magnetic flux transport, we do not intend to evolve
the system for much longer, nor to address the evolutionary effects in further
detail. We simply note here that the rate of flux transport we have measured
should only be considered as a reference, and can be subject to uncertainties
associated with global conditions.

\bibliographystyle{apj}

\label{lastpage}
\end{document}